\documentstyle[amsfonts,aps]{revtex}

\begin{document}
\author{Mario Castagnino}
\address{Instituto de Astronom\'{\i}a y F\'{\i}sica del Espacio,\\
Casilla de Correos 67, Sucursal 28,\\
1428 Buenos Aires, Argentina.}
\author{Edgard Gunzig}
\address{Instituts Internationaux de Physique et de Chimie,\\
50 Av. F. D. Roosevelt,\\
1050 Bruxelles, Belgium.}
\title{Minimal Irreversible Quantum Mechanics: An Axiomatic Formalism. }
\maketitle

\begin{abstract}
An axiomatic formalism for a minimal irreversible quantum mechanics is
introduced. It is shown that a quantum equilibrium and the decoherence
phenomenon are consequences of the axioms and that Lyapunov variables,
exponential survival probabilities, and a classisal conditional
never-decreasing entropy can be defined.

\begin{itemize}
\item  e-mail: castagni(a)iafe.uba.ar

\item  Pacs Nrs.: 03.65.Bz, 05.45+b, 05.70.Ln
\end{itemize}
\end{abstract}

\section{ Introduction.}

Let us consider the function $y=f(x)=x^2$. Can we say if this function is an
even function or an odd function? The primary (but incorrect) answer would
be that it is an even function. This answer is wrong because, in order to
define a function properly, we must also define its domain of definition $D$
and its range $R$, namely: 
\[
f(x)=y 
\]
\begin{equation}
f:D\rightarrow R  \label{1.1}
\end{equation}
Then if $y=f(x)=x^2$ is defined as $f:{\Bbb R}{\bf \rightarrow }{\Bbb R}_{+%
\text{ }}$it is an even function. But if it is defined as $f:{\Bbb R}%
_{+}\rightarrow {\Bbb R}_{+}$ the function is neither even nor odd. The
morale of this story is that when we speak about a symmetry of a function
necessarily we must define its domain of definition and its range, if not
what we may say could be meaningless.

Let us now consider the Schroedinger equation: 
\begin{equation}
i\frac{d|\psi >}{dt}=H|\psi >  \label{1.2}
\end{equation}
and its solution: 
\begin{equation}
|\psi (t)>=e^{-iHt}|\psi (0)>  \label{1.3}
\end{equation}
From what we have just learnt the question: ''is the set of the time
evolutions obtained from Schroedinger equation time-reversible or
invertible? \cite{Messiah}, \cite{CGG}'' {\bf has no meaning} if we do not
define the domain of definition and the range of the states $|\psi >$ in eq.
(\ref{1.3}), i. e. the space where this vectors live. If we choose a Hilbert
space ${\cal H}$ the set of time-evolutions of eq. (\ref{1.3}) is
time-symmetric and each evolution is invertible. But if we make a different
choice the set can become time asymmetric and each time-evolution can become
non-invertible. Let us explain why it is so.

Let $K$ be the Wigner time-inversion operator. Hilbert space is invariant
under time-inversion, namely: 
\begin{equation}
K:{\cal H\rightarrow H}  \label{1.4}
\end{equation}
But we can choose a non-time-reversal-invariant space as the space of
physical admissible state, let be $\phi _{-\text{ }},$ such that: 
\begin{equation}
K:\phi _{-}\rightarrow \phi _{+}\neq \phi _{-}  \label{1.5}
\end{equation}
and then within this space the set of time evolutions will turn out to be
time-asymmetric and each evolution non invertible, as it is shown in the
literature (precisely in almost all bibliographical references of this
paper) and as we will also demonstrate below.

In this minimal way we can obtain a natural irreversible quantum mechanics.
The aim of this paper is to sketch, using the results of many authors quoted
in the references and our own results, an axiomatic formalism for this
theory, which may have two possible advantages over ordinary quantum
mechanics:

i.- Universe is clearly time asymmetric. Then the new theory may describe
the real universe better than the usual one. We shall further discuss this
possibility in section 16.

ii.- The new theory has more powerful spectral decompositions that makes the
study of decaying processes easier

Let us rephrase what we have said using physical language: If we forget the
time-asymmetric weak interaction (as it is usual in this kind of research,
since weak interaction is so weak that it is difficult to see how it can
explain the macroscopic time-asymmetry \cite{Sachs}), the time-asymmetry
problem can be stated in the following question:

{\it How can we explain the obvious time-asymmetry of the universe and most
of its subsystems if the fundamental laws of physic are time-symmetric?}

There are only two causes for asymmetry in nature: either the laws of nature
are asymmetric or the solutions of the equations of the theory are
asymmetric. As time asymmetry is not an exception the answer is contained in
the question itself: If the laws of nature are time-symmetric essentially
the only way we have to explain the time-asymmetry of the universe is to
postulate that the state of the universe, or more generally, the space of
physical possible solutions of the universe evolution equations is not
time-reversal invariant, namely to use the second cause of asymmetry \cite
{CGG}, \cite{Feynman} . In this paper we explore this possibility using an
axiomatic framework.

Moreover, certainly the best way to explain a physical idea is to construct
an axiomatic structure because, having this structure, somehow we can see
the whole idea, even if we cannot foresee all its consequences. Analogously
it is easier to criticize an idea when it is presented in an axiomatic
language. So we believe that this paper can clarify some issues of the
problem of time-asymmetry.

The paper is organized as follows: Section 2: we define the space and the
notation we will use. Section 3: the analytic continuation of the solutions
is studied. Section 4: density matrices and Liouville space are introduced.
Section 5: the space for the observables is chosen. Section 6: the axioms of
the theory are stated. Sections 7 and 8: the main consequences of the axioms
are obtained. Section 9: time-asymmetry and irreversibility are studied.
Section 10: it is shown how Schr\"{o}dinger and Heisenberg pictures work in
the new formalism. Section 11: quantum equilibrium and decoherence are
obtained. Section 12: it is shown that the norm and the energy are conserved
and how Lyapunov variables appear. Section 13: entropy is defined. Section
14: the thermalization phenomenon is studied. Section 15: the global nature
of the time-asymmetry is considered. Section 16: the Reichenbach diagram is
presented. Section 17: other results are listed. Section 18: we draw our
main conclusions. An appendix completes the paper.

\section{Definition space.}

Let us consider a quantum system, with a free hamiltonian $H_0$, endowed
with a discrete plus a continuous spectrum, namely such that: 
\[
H_0|E_n^{(0)}>=E_n^{(0)}|E_n^{(0)}> 
\]
\begin{equation}
H_0|E^{(0)}>=E^{(0)}|E^{(0)}>  \label{2.1}
\end{equation}
where, $n=(0,1,2,...N_0),$ $0\leq E^{(0)}\leq \infty ,$ $E_n^{(0)}\geq 0.$
The total hamiltonian will be $H=H_0+V,$ and the perturbation will be such
that some bound states of the discrete spectrum become complex poles, namely
we will have: 
\[
H|E_n>=E_n|E_n> 
\]
\begin{equation}
H|E>=E|E>  \label{2.1'}
\end{equation}
where, $n=(0,1,2,...N),$ $0\leq E\leq \infty ,$ and $N<N_0$ (in almost all
the cases, for simplicity and in order to fix the ideas, we will consider
that $N=0$, and therefore there is only one discrete {\it ground state}; a
more general case will be considered in section 14)$.\{|E_n>,E\pm >\}$ is a
basis of the corresponding Hilbert space ${\cal H.}$ (e.g. $|E\pm >$ can be
the Lippmann-Schwinger retarded or advanced bases $\{|\omega _{\pm }>\}$ of
ref. \cite{Bohm}), and: 
\begin{equation}
<E_n|E_m>=\delta _{nm},\text{ }<E\pm |E^{\prime }\pm >=\delta (E-E^{\prime
}),\text{ }<E\pm |E_n>=0  \label{2.2}
\end{equation}
\begin{equation}
I=\sum_{n=0}^{n=N}|E_n><E_n|+\int_0^\infty |E\pm ><E\pm |dE  \label{2.3}
\end{equation}
\begin{equation}
H=\sum_{n=0}^{n=N}E_n|E_n><E_n|+\int_0^\infty E|E\pm ><E\pm |dE  \label{2.3'}
\end{equation}
Let $\Xi $ be the vector space of all possible linear combinations of the
basis $\{|E_n>,|E\pm >\}$ vectors \cite{Ballentine} so if $|\psi >\in \Xi :$%
\begin{equation}
|\psi >=\sum_{n=1}^{n=N}\psi _n|E_n>+\int_0^\infty \psi _{\pm }(E)|E\pm >dE
\label{2.4}
\end{equation}
where neither $\psi _n$ nor $\psi (E)$ have any peculiar property.

Let $K$ be the Wigner time inversion operator \cite{Messiah} therefore: 
\begin{equation}
K|E_n>=|E_n>,\text{ }K|E\pm >=|E\mp >  \label{2.5}
\end{equation}
(for the continuous spectrum the Lippmann-Schwinger advanced and retarded
bases has this property). Then, as $K$ is antilinear: 
\begin{equation}
K|\psi >=\sum_{n=1}^{n=N}\psi _n^{*}|E_n>+\int_0^\infty \psi _{\pm
}^{*}(E)|E\mp >dE  \label{2.6}
\end{equation}
To find an irreversible quantum mechanics we must define a subspace $\phi
_{-}$ of $\Xi $ such that: 
\begin{equation}
K:\phi _{-}\rightarrow \phi _{+}\neq \phi _{-}  \label{2.7}
\end{equation}
namely a subspace which is not invariant under time inversions.

In our opinion nowadays there is a unique way to define space $\phi _{-}$
(see \cite{L&P}, \cite{AntoniouBohm}, \cite{Bohm51}, \cite{mex}). In fact it
is completely reasonable to ask that $\phi _{-}$ would have some logical
properties namely that $\phi _{-}\subset {\cal H}$ (i.e. $\psi _n\in l^2,$ $%
\psi _{\pm }(E)\in L^2[0,\infty )$), $\phi _{-}$ must be dense in ${\cal H}%
_{-}$ (the outgoing state subspace of ${\cal H}$) and its topology must be a
nuclear one. Precisely we will define the spaces ${\cal H}_{-}$ and $\phi
_{-}$ as: 
\[
|\psi >\in {\cal H}_{-}\Longleftrightarrow \psi _{+}(E)\in H_{-}^2|_{{\Bbb R}%
_{+}} 
\]
\begin{equation}
|\psi >\in \phi _{-}\Longleftrightarrow \psi _{+}(E)\in {\cal S}\cap
H_{-}^2|_{{\Bbb R}_{+}}={\cal S}_{-}  \label{2.8}
\end{equation}
where ${\cal S}$ is the Schwarz class function space (this choice allows to
perform derivative to any order) and $H_{-}^2$ is the space of Hardy class
function from below \cite{Bohm} (this choice introduces causality in our
theory \cite{mex}). Nevertheless other choices have being used: \cite{Diener}%
, \cite{Gadella}.

As $\phi _{-}\subset {\cal H}_{-}$ we have the Gel'fand triplet: 
\begin{equation}
\phi _{-}\subset {\cal H}_{-}\subset \phi _{-}^{\times }  \label{2.9}
\end{equation}
where $\phi _{-}^{\times }$ is the space of antilinear functionals $F$ over $%
\phi _{-}$, such that : 
\begin{equation}
F[\psi ]=<\psi |F>=<F|\psi >^{*}  \label{2.9'}
\end{equation}
This will be the main arena of all our calculations. But, as we will see, we
must also use the time inverted objects. Precisely, the spaces ${\cal H}_{+}$
and $\phi _{+}$ defined as: 
\[
|\psi >\in {\cal H}_{+}\Longleftrightarrow \psi _{-}(E)\in H_{+}^2|_{{\Bbb R}%
_{+}} 
\]

\begin{equation}
|\psi >\in \phi _{+}\Longleftrightarrow \psi _{-}(E)\in {\cal S}\cap
H_{+}^2|_{{\Bbb R}_{+}}={\cal S}_{+}  \label{2.10}
\end{equation}
where $H_{+}^2$ is the Hardy class from above, and the Gel'fand triplet is:

\begin{equation}
\phi _{+}\subset {\cal H}_{+}\subset \phi _{+}^{\times }  \label{2.11}
\end{equation}
It is easy to see that the spaces $\phi _{-}$ and $\phi _{+}$ satisfy eq. (%
\ref{2.7}).

We close the section with three observations.

i.- Hamiltonian $H$ must be time-independent, since our aim is to define an
arrow of time in a closed system. In fact, a realistic arrow of time must be
defined using the whole universe as the system \cite{Castscape.} (open
systems will be considered in section 14).

ii.- At first sight we can think that with the method we are about to
propose we can define time asymmetry in non-interacting system like a free
particle. It is not so since, even if the resulting free particle theory
would formally be time-asymmetric, the entropy will not grow. In fact, the
entropy will only grow, as we will see, if we have a non-trivial S-matrix
with complex poles, which is not the case of a trivial free particle.

iii.- $\phi _{-}$ is dense in ${\cal H}_{-}$ so, if someone would say that
the ''real'' physical states are those of ${\cal H}_{-}{\cal ,}$ we can
answer that any one of these states can be approximated, as close as we
wish, with a state of $\phi _{-}.$ So, on physical measurement grounds, the
states of both spaces are indistinguishable. Nevertheless the two spaces
have different kinds of topologies.

\section{Analytic continuations.}

Let us consider an scattering experiment using hamiltonian $H$ and let $%
\{|\omega _{+}>\}$ be the Lippmann-Schwinger basis (all objects related with
this basis will be labelled with $\omega ,$ instead of the $E$ used in the
equations of the last section), then we know that: 
\begin{equation}
\sum_{n=0}^N|\omega _n><\omega _n|+\int_0^\infty |\omega _{+}><\omega
_{+}|d\omega =I  \label{L-S}
\end{equation}
where the $|\omega _n>$ are the eventual stable bound states. Then: 
\begin{equation}
<\varphi |\psi >=\sum_{n=0}^N<\varphi |\omega _n><\omega _n|\psi
>+\int_0^\infty <\varphi |\omega _{+}><\omega _{+}|\psi >d\omega
\label{L-S-1}
\end{equation}

Let $z_{n\text{ }}$be the real and complex poles of the corresponding
S-matrix. Then, using a simple analytic continuation of eq. (\ref{L-S-1}) it
can be demonstrated \cite{Bohm} that if $|\psi >\in \phi _{-}$ and $|\varphi
>\in \phi _{+},$ the inner product $<\varphi |\psi >$ (which is well defined
since both vectors belong to ${\cal H}$ ) reads: 
\begin{equation}
<\varphi |\psi >=\sum_{n=0}^{N_0}<\varphi |\overline{f_n}><\widetilde{f_n}%
|\psi >+\int_\Gamma <\varphi |\overline{f_z}><\widetilde{f_z}|\psi >dz
\label{3.0}
\end{equation}
where $\Gamma $ is a curve that begins at $O$ and goes to $+\infty $ of the
real axis under all the poles of the lower halfplane. Also, making the
Nakanishi trick \cite{CGG}, \cite{Nakanishi} we can obtain: 
\begin{equation}
<\varphi |\psi >=\sum_{n=0}^{N_0}<\varphi |\overline{f_n}><\widetilde{f_n}%
|\psi >+\int_0^\infty <\varphi |\overline{f_\omega }><\widetilde{f_\omega }%
|\psi >d\omega  \label{3.1}
\end{equation}
where there is a term in the sum for each pole of the S-matrix, precisely:
for each complex poles and each real pole corresponding to the bound states
of the sum of eq. (\ref{L-S}). Analogously it can be demonstrated that:

\begin{equation}
<\varphi |H|\psi >=\sum_{n=0}^{N_0}z_n<\varphi |\overline{f_n}><\widetilde{%
f_n}|\psi >+\int_0^\infty \omega <\varphi |\overline{f_\omega }><\widetilde{%
f_\omega }|\psi >d\omega  \label{3.2}
\end{equation}
(se also \cite{Trio}) where $|\overline{f_n}>,|\overline{f_\omega }>\in \phi
_{+}^{\times },$ $|\widetilde{f_n}>,|\widetilde{f_\omega }>\in \phi
_{-}^{\times }$ and in particular $|\widetilde{f_n}>=|\omega _n>$ (if $0\leq
n\leq N)$ and $|\widetilde{f}_\omega >=|\omega _{+}>$ (eq. (44) ref. \cite
{Laucast}).

Also if:

$z_n$ is real it is the eigenenergy of a bound state and if,

$z_n$ is complex it is a pole of the S-matrix

Therefore any $|\psi >\in \phi _{-}$ reads

\begin{equation}
|\psi >=\sum_{n=0}^{N_0}|\overline{f_n}><\widetilde{f_n}|\psi
>+\int_0^\infty |\overline{f_\omega }><\widetilde{f_\omega }|\psi >d\omega
\label{3.3}
\end{equation}
in a weak sense (namely premultiplied by any $<\varphi |\in \phi _{+})$ and $%
<\widetilde{f_\omega }|\psi >\in {\cal S}_{-}$

Analogously: 
\begin{equation}
H|\psi >=\sum_{n=0}^{N_0}z|\overline{f_n}><\widetilde{f_n}|\psi
>+\int_0^\infty \omega |\overline{f_\omega }><\widetilde{f_\omega }|\psi
>d\omega  \label{3.A}
\end{equation}
Then, in an even weaker sense the two last equations can be written as: 
\begin{equation}
I=\sum_{n=0}^{N_0}|\overline{f_n}><\widetilde{f_n}|+\int_0^\infty |\overline{%
f_\omega }><\widetilde{f_\omega }|d\omega  \label{3.B}
\end{equation}
\begin{equation}
H=\sum_{n=0}^{N_0}z|\overline{f_n}><\widetilde{f_n}|+\int_0^\infty \omega |%
\overline{f_\omega }><\widetilde{f_\omega }|d\omega  \label{3.C}
\end{equation}
The bases $\{|\overline{f_n}>,|\overline{f_\omega }>\}$ and $\{|\widetilde{%
f_n}>,|\widetilde{f_\omega }>\}$ are a biorthonormal system \cite{Bohm}, 
\cite{Sudarshan} namely: 
\begin{equation}
<\widetilde{f_n}|\overline{f_m}>=\delta _{nm},\text{ }<\widetilde{f_n}|%
\overline{f_\omega }>=0,\text{ }<\widetilde{f_\omega }|\overline{f_n}>=0,%
\text{ }<\widetilde{f_\omega }|\overline{f_{\omega ^{\prime }}}>=\delta
(\omega -\omega ^{\prime })  \label{3.D}
\end{equation}
Also it can be proved that: 
\begin{equation}
<\overline{f_n}|\overline{f_m}>=\delta _{nm}\varepsilon _n  \label{3.E}
\end{equation}
\begin{equation}
<\widetilde{f_n}|\widetilde{f_m}>=\delta _{nm}\varepsilon _n  \label{3.F}
\end{equation}

where $\varepsilon _n=1$ if $%
\mathop{\rm Im}%
z_n=0$ and $\varepsilon _n=0$ otherwise \cite{Laucast}, \cite{Gadella},
Namely the states with $%
\mathop{\rm Im}%
z_n\neq 0$ are ''ghosts'' with vanishing norm. This fact is evident since if 
$|\overline{f_n}>$ is one of these ghosts, from eq. (\ref{L-S}), we have: 
\begin{equation}
<\overline{f_n}|\overline{f_n}>=<\overline{f_n}|(\sum_{n=0}^N|\omega
_n><\omega _n|+\int_0^\infty |\omega _{+}><\omega _{+}|d\omega )|\overline{%
f_n}>=  \label{3.10}
\end{equation}
\[
=<\overline{f_n}|(\sum_{n=0}^N|\widetilde{f_n}><\widetilde{f_n}%
|+\int_0^\infty |\widetilde{f}_\omega ><\widetilde{f}_\omega |d\omega )|%
\overline{f_n}>=0 
\]
where we have used eq. (\ref{3.B}) and that $|\widetilde{f_n}>=|\omega _n>$%
and $|\widetilde{f}_\omega >=|\omega _{+}>$ (eq. (44) of ref. \cite{Laucast}%
).

We will, sometimes, find useful write all these equation using a shorthand
notation where we will call the basis $\{|\omega _0>,|\omega +>\}$ just $%
\{|i>\},$ the basis $\{|\overline{f_n}>,|\overline{f_\omega }>\}$ just $\{|%
\overline{i}>\},$ and the basis $\{|\widetilde{f_n}>,|\widetilde{f_\omega }%
>\}$ just$|\{|\widetilde{i}>\}.$ Then eq. (\ref{3.3}) reads: 
\begin{equation}
|\psi >=\sum_i|\overline{i}><\widetilde{i}|\psi >=\sum_i\psi _i|\overline{i}>
\label{3.4}
\end{equation}
and also we will conventionally say that $<\widetilde{i}|\psi >\in {\cal S}%
_{-}.$ Eq. (\ref{3.A}) reads: 
\begin{equation}
H|\psi >=\sum_iz_i|\overline{i}><\widetilde{i}|\psi >  \label{3.5}
\end{equation}
In all these equations $|\overline{i}>\in \phi _{+}^{\times },|\widetilde{i}%
>\in \phi _{-}^{\times }.$ The biorthonormality of the system $\{|\overline{i%
}>\}$ and $\{|\widetilde{i}>\}$ will be symbolized as: 
\begin{equation}
<\widetilde{i}|\overline{j}>=\delta _{ij}  \label{3.6}
\end{equation}
\begin{equation}
\sum_i|\overline{i}><\widetilde{i}|=I  \label{3.7}
\end{equation}
where the symbols have an obvious meaning (e. g. eq. (\ref{3.7}) is a
shorthand-notation weak version of eq. (\ref{3.B}), etc.).

Also: 
\begin{equation}
<\overline{i}|\overline{j}>=\delta _{ij}\varepsilon _i  \label{3.8}
\end{equation}
\begin{equation}
<\widetilde{i}|\widetilde{j}>=\delta _{ij}\varepsilon _i  \label{3.9}
\end{equation}

where $\varepsilon _i=1$ if $%
\mathop{\rm Im}%
z_i=0$ and $\varepsilon _i=0$ in all the other cases.

\section{Density matrices.}

Up to now we have just introduced pure states, but we can rephrase
everything using mixed states $\rho $. In general $\rho \in \Xi \otimes \Xi $
but usually it is considered that it belongs to a Liouville space ${\cal L}%
^{\prime }{\cal =H\otimes H.}$ The time evolution of the mixed states can be
obtained solving Liouville equation: 
\begin{equation}
i\frac{d\rho }{dt}=[H,\rho (t)]=L\rho (t)  \label{Liouville}
\end{equation}
where $L$ is the Liouville operator.

From this equation we see that any $\rho _{*}$ that commutes with $H$ is an
stationary state. This state is diagonal in the same basis than $H$ and
therefore it can be written as (cf. eq. (\ref{2.3'})): 
\begin{equation}
\rho _{*}=\rho _0|\omega _0><\omega _0|+\int_0^\infty \rho _\omega |\omega
+><\omega +|d\omega  \label{4.A}
\end{equation}
where we have made $N=0$ for simplicity as we have announced. The second
term of the r. h. s. of the last equation implies the existence of a
singular structure in the stationary state, that was studied at large in the
paper \cite{LaucastIII}. So, if we want to develop a rigorous treatment of
this singular structure we are forced to consider that ${\cal L=H\oplus
(H\otimes H)}$, where the first ${\cal H}$ contains the singular structure
and the second factor ${\cal H\otimes H}$ is the usual Liouville space $%
{\cal L}^{\prime },$ that now it will be only considered as a regular
structure, and therefore we introduce the following eigenbasis of $L$: 
\[
\rho (0)=|\omega _0><\omega _0|,\text{ }\rho (0,\omega )=|\omega _0><\omega
+|,\text{ }\rho (\omega ,0)=|\omega +><\omega _0| 
\]
\begin{equation}
\beta (\omega )=|\omega +><\omega +|,\text{ }\rho (\omega ,\omega ^{\prime
})=|\omega +><\omega ^{\prime }+|  \label{4.B}
\end{equation}
This is an orthonormal basis in an inner product that we will define below
(cf. eq. (\ref{4.6})).

We can now compute the eigenvalues of the eigenvalues of $L:$%
\[
L\rho (0)=0,\text{ }L\rho (0,\omega )=(\omega _0-\omega )\rho (0,\omega ),%
\text{ }L\rho (\omega ,0)=(\omega -\omega _0)\rho (\omega ,0) 
\]
\begin{equation}
L\beta (\omega )=0,\text{ }L\rho (\omega ,\omega ^{\prime })=(\omega -\omega
)\rho (\omega ,\omega ^{\prime })  \label{4.C}
\end{equation}

But for $\rho (\omega ,\omega ^{\prime })$ it is better to use Riesz quantum
numbers: 
\[
\sigma =\frac 12(\omega +\omega ^{\prime }),\text{ }0\leq \sigma <\infty 
\]
\begin{equation}
\nu =\omega -\omega ^{\prime },\text{ }-2\sigma \leq \nu \leq 2\sigma
\label{4.4}
\end{equation}
So we will write the matrices $\rho (\omega ,\omega ^{\prime })$ as: 
\begin{equation}
\rho (\omega ,\omega ^{\prime })=\beta (\sigma ,\nu )  \label{4.5}
\end{equation}
So any $\rho \in {\cal L}$ can be written as:

\begin{equation}
\rho =\rho _0\rho (0)+\int_0^\infty [\rho _{0\omega }\rho (0,\omega )+\rho
_{\omega 0}\rho (\omega ,0)+\rho _\omega \beta (\omega )]d\omega
+\int_0^\infty d\sigma \int_{-2\sigma }^{2\sigma }d\nu \rho _{\sigma \nu
}\beta (\sigma ,\nu )  \label{4.9}
\end{equation}
The inner product among these $\rho $ is naturally defined as: 
\begin{equation}
(\rho |\rho ^{\prime })=\rho _0^{*}\rho _0^{\prime }+\int_0^\infty [\rho
_{0\omega }^{*}\rho _{0\omega }^{\prime }+\rho _{\omega 0}^{*}\rho _{\omega
0}^{\prime }+\rho _\omega ^{*}\rho _\omega ^{\prime }]d\omega +\int_0^\infty
d\sigma \int_{-2\sigma }^{2\sigma }d\nu \rho _{\sigma \nu }^{*}\rho _{\sigma
\nu }^{\prime }  \label{4.6}
\end{equation}
From eq. (\ref{4.C}), we have. 
\begin{equation}
L=\int_0^\infty (\omega _0-\omega )[\rho (0,\omega )\rho ^{\dagger
}(0,\omega )-\rho (\omega ,0)\rho ^{\dagger }(\omega ,0)]d\omega
+\int_0^\infty d\sigma \int_{-2\sigma }^{2\sigma }\nu d\nu \beta (\sigma
,\nu )\beta ^{\dagger }(\sigma ,\nu )  \label{4.7}
\end{equation}
Now let us make the analytic continuation.:

The diagonal elements $\rho (0)$ and $\beta (\omega )$ will remain
untouched, since they correspond to the stationary state, but we will
require that : 
\begin{equation}
\rho _\omega \in {\cal S.}  \label{4.7'}
\end{equation}

The terms $\rho (0,\omega )$ and $\rho (\omega ,0)$ can be treated as in the
last section, so they have only one variable $\omega ,$ so we will ask that: 
\begin{equation}
\rho _{\omega 0}\in {\cal S}\cap H_{-}^2|_{{\Bbb R}_{+}}={\cal S}_{-},\text{ 
}\rho _{0\omega }\in {\cal S}\cap H_{+}^2|_{{\Bbb R}_{+}}={\cal S}_{+}
\label{4.8}
\end{equation}

Finally let us consider the term $\beta (\sigma ,\nu ).$ We could promote
both real variables $\sigma $ and $\nu $ to complex variables but, as $\nu $
is the eigenvalue of the Liouville operator, it is only necessary to promote 
$\nu \rightarrow z\in {\Bbb C}$ \cite{Sudarshanu} and leave $\sigma $ real.
Precisely as: 
\begin{equation}
\rho _{\sigma \nu }=(\beta (\sigma ,\nu )|\rho )  \label{4.10}
\end{equation}
we can consider the complex value function of $z:$%
\begin{equation}
\rho _{\sigma z}=(\beta (\sigma ,z)|\rho )  \label{4.11}
\end{equation}
and to ask that: 
\begin{equation}
\rho _{\sigma \nu }\in {\cal S}\cap H_{-}^2|_{-2\sigma }^{2\sigma }={\cal S}%
_{-}^{(\sigma )}  \label{4.12}
\end{equation}
for any $\sigma \geq 0,$ thus $\rho _{\sigma z}$ with be an analytical
function of $z$ in the lower halfplane. Then we will say that $\rho \in \Phi
_{-}$ if eqs. (\ref{4.7'}), (\ref{4.8}), and (\ref{4.12}) are $\rho _{\omega
0}\in {\cal S}\cap H_{-}^2|_{{\Bbb R}_{+}}={\cal S}_{-},$ $\rho _{0\omega
}\in {\cal S}\cap H_{+}^2|_{{\Bbb R}_{+}}={\cal S}_{+}$. We also define a
space ${\cal L}_{-}$ such that if $\rho \in {\cal L}_{-}$ we simply have : 
\[
\rho _{\omega 0}\in H_{-}^2|_{{\Bbb R}_{+}},\text{ }\rho _{0\omega }\in
H_{+}^2|_{{\Bbb R}_{+}} 
\]
\[
\rho _{\sigma \nu }\in H_{-}^2|_{-2\sigma }^{2\sigma } 
\]
Let us now define the time-inverted spaces $\Phi _{+}$ and ${\cal L}_{+}.$
If eq. (\ref{4.7'}) is satisfied and : 
\begin{equation}
\rho _{\omega 0}\in {\cal S}\cap H_{+}^2|_{{\Bbb R}_{+}}={\cal S}_{+},\text{ 
}\rho _{0\omega }\in {\cal S}\cap H_{-}^2|_{{\Bbb R}_{+}}={\cal S}_{-}
\label{4.12'}
\end{equation}
\begin{equation}
\rho _{\sigma \nu }\in {\cal S}\cap H_{+}^2|_{-2\sigma }^{2\sigma }={\cal S}%
_{+}^{(\sigma )}  \label{4.13}
\end{equation}
for any $\sigma \geq 0,$ thus in this case $\rho _{\sigma z}$ will be an
analytical function of $z$ in the upper halfplane, then we will say that $%
\rho \in \Phi _{+}$ (but in the definition of this space the basis $|\omega
+>$ of eq. (\ref{4.B}) must be changed by the basis $|\omega ->).$

We also define a space ${\cal L}_{+}$ such that if $\rho \in {\cal L}_{+}$
we have: 
\[
\rho _{\omega 0}\in H_{+}^2|_{{\Bbb R}_{+}},\text{ }\rho _{0\omega }\in
H_{-}^2|_{{\Bbb R}_{+}} 
\]
\[
\rho _{\sigma \nu }\in H_{+}^2|_{-2\sigma }^{2\sigma } 
\]
where we have also changed the basis as before.

Let us now consider the poles:

For the terms $\rho (0,\omega )$ and $\rho (\omega ,0)$ we will find those
of the last section, and we can repeat the analytic continuation up to the
curve $\Gamma $ of section 3.

For the terms $\beta (\omega )$ and $\beta (\sigma ,\nu )$, for some fixed $%
\sigma $ and for every pole $z_n$ of the S-matrix we will find at the $\nu $
or $z-$plane two poles $\pm 2(z_n-\sigma )$ (and also a pole at $\nu =z=0$
coming from the singular structure of the continuous field, the $\beta
(\omega )$ term)$,$ that we will call $\zeta _l.$ Also it may happen that
for some $\sigma _j$ extra poles $\zeta _l^j$ may appear \cite{Roberto}. So
introducing a curve $C,$ in the lower halfplane, that goes, under all the
poles, from $-2\sigma $ to $2\sigma $ of the real axis (fig. 1) and using,
as in the pure states case, the Cauchy theorem, if $\rho \in \Phi _{+}$ and $%
\rho ^{\prime }\in \Phi _{-},$ we obtain that: 
\[
(\rho |\rho ^{\prime })=(\rho |\{\rho (0)\rho ^{\dagger }(0)+ 
\]
\[
+\sum_n[\rho (\overline{z_{n,}0})\rho ^{\dagger }(\widetilde{z_n,0)}+\rho (%
\overline{0,z_n})\rho ^{\dagger }(\widetilde{0,z_n})]+\int_\Gamma [\rho (%
\overline{z,0})\rho ^{\dagger }(\widetilde{z,0})+\rho (\overline{0,z})\rho
^{\dagger }(\widetilde{0,z})]dz+ 
\]
\begin{equation}
\sum_{jl}\overline{\beta (\sigma _{j,}\zeta _l^j)}\widetilde{\beta ^{\dagger
}(\sigma _{j,}\zeta _l^j)}+\int_0^\infty d\sigma [\beta (\sigma )\beta
^{\dagger }(\sigma )+\sum_l\beta (\overline{\sigma ,\zeta _l})\beta
^{\dagger }\widetilde{(\sigma ,\zeta _l)}+\int_C\beta (\overline{\sigma z}%
)\beta ^{\dagger }(\widetilde{\sigma z})dz]\}|\rho ^{\prime })  \label{4.D}
\end{equation}
Namely in weak sense: 
\[
I=\rho (0)\rho ^{\dagger }(0)+\sum_n[\rho (\overline{z_{n,}0})\rho ^{\dagger
}(\widetilde{z_n,0})+\rho (\overline{0,z_n})\rho ^{\dagger }(\widetilde{0,z_n%
})+\int_\Gamma [\rho (\overline{z,0})\rho ^{\dagger }(\widetilde{z,0})+\rho (%
\overline{0,z})\rho ^{\dagger }(\widetilde{0,z})]dz+ 
\]
\begin{equation}
\sum_{jl}\overline{\beta (\sigma _{j,}\zeta _l^j)}\widetilde{\beta ^{\dagger
}(\sigma _{j,}\zeta _l^j)}+\int_0^\infty d\sigma [\beta (\sigma )\beta
^{\dagger }(\sigma )+\sum_l\beta (\overline{\sigma ,\zeta _l})\beta
^{\dagger }\widetilde{(\sigma ,\zeta _l})+\int_C\beta (\overline{\sigma z}%
)\beta ^{\dagger }(\widetilde{\sigma z})dz]  \label{4.E}
\end{equation}
In these equations the presence of the poles coming from the singular
structure ( in each $\sigma =const.$ plane) is represented by the terms $%
\beta (\sigma )\beta ^{\dagger }(\sigma ).$

Then we can write any $\rho \in \Phi _{-}$ as: 
\[
\rho =\rho _0\rho (0)+\sum_n[\rho _{n0}\rho (\overline{z_{n,}0})+\rho
_{0n}\rho (\overline{0,z_n})+\int_\Gamma [\rho _{z0}\rho (\overline{z,0}%
)+\rho _{0z}\rho (\overline{0,z})]dz 
\]
\begin{equation}
\sum_{jl}\overline{\rho _{jl}\beta (\sigma _{j,}\zeta _l^j)}+\int_0^\infty
d\sigma [\rho _\sigma \beta (\sigma )+\sum_l\rho _{\sigma l}\beta (\overline{%
\sigma ,\zeta _l})+\int_C\rho _{\sigma z}\beta (\overline{\sigma z})dz]
\label{4.F}
\end{equation}

Analogously, from the analytic continuation of the Liouville operator we
obtain: 
\[
(\rho |L|\rho ^{\prime })=(\rho |\{\sum_n[(z_n-\omega _0)\rho (\overline{%
z_{n,}0})\rho ^{\dagger }(\widetilde{z_n,0)}+(\omega _0-z_{n)}^{*}\rho (%
\overline{0,z_n})\rho ^{\dagger }(0,z_n)]+ 
\]
\[
+\int_\Gamma [(z-\omega _0)\rho (\overline{z,0})\rho ^{\dagger }(\widetilde{%
z,0})+(\omega _0-z^{*})\rho (\overline{0,z})\rho ^{\dagger }(\widetilde{0,z}%
)]dz+ 
\]
\begin{equation}
\sum_{jl}\overline{\beta (\sigma _{j,}\zeta _l^j)}\widetilde{\beta ^{\dagger
}(\sigma _{j,}\zeta _l^j)}+\int_0^\infty d\sigma [\sum_l\zeta _l\beta (%
\overline{\sigma ,\zeta _l})\beta ^{\dagger }(\widetilde{\sigma ,\zeta _l}%
)|\rho ^{\prime })+\int_Cz\beta (\overline{\sigma ,z})\beta ^{\dagger }(%
\widetilde{\sigma ,z})|\rho ^{\prime })dz]\}|\rho ^{\prime })  \label{4.14}
\end{equation}
where $\zeta _l=\overline{\nu }_l-i\gamma _l,$ $\gamma _l\geq 0$ and $\zeta
_l^j=\overline{\nu }_l^j-i\gamma _l^j,$ $\gamma _l^j\geq 0$, and as in the
pure states case, $\Gamma $ is a curve that goes from $0$ to $+\infty $ of
the real axis under all the poles of the lower half plane. Namely, in weak
sense, the Liouville operator reads: 
\[
L=\{\sum_n[(z_n-\omega _0)\rho (\overline{z_{n,}0})\rho ^{\dagger }(%
\widetilde{z_{n,}0})+(\omega _0-z_{n)}^{*}\rho (\overline{0,z_n})\rho
^{\dagger }(0,z_n)]+ 
\]
\[
+\int_\Gamma [(z-\omega _0)\rho (\overline{z,0})\rho ^{\dagger }(\widetilde{%
z,0})+(\omega _0-z^{*})\rho (\overline{0,z})\rho ^{\dagger }(\widetilde{0,z}%
)]dz+ 
\]
\begin{equation}
\sum_{jl}\zeta _l^j\overline{\beta (\sigma _{j,}\zeta _l^j)}\widetilde{\beta
^{\dagger }(\sigma _{j,}\zeta _l^j)}+\int_0^\infty d\sigma [\sum_l\zeta _l%
\overline{\beta (\sigma ,\zeta _l})\beta ^{\dagger }(\widetilde{\sigma
,\zeta _l})+\int_{-2\sigma }^{2\sigma }\nu \beta (\overline{\sigma ,\nu }%
)\beta ^{\dagger }(\widetilde{\sigma .\nu })d\nu ]  \label{4.15}
\end{equation}
As in the pure states case the bases:

$\{\rho (0),\rho (\overline{0,z_n}),\rho (\overline{z_n,0}),\rho (\overline{%
0,z}),\rho (\overline{z,0}),\beta (\omega ),\overline{\beta (\sigma
_{j,}\zeta _l^j)},\beta (\overline{\sigma ,\zeta _l}),\beta (\overline{%
\sigma ,z})\}$ and

$\{\rho (0),\rho (\widetilde{0,z_n}),\rho (\widetilde{z_n,0}),\rho (%
\widetilde{0,z}),\rho (\widetilde{z,0}),\beta (\omega ),\widetilde{\beta
(\sigma _j,\zeta _l^j),}\beta (\widetilde{\sigma ,\zeta _l}),\beta (%
\widetilde{\sigma ,z})\}$

are a biorthonormal system under the inner product (\ref{4.6}) \footnote{%
From now on we will consider that the discrete index $\sigma _j$ is included
in the continuous index $\sigma ,$ and also $\zeta _l^j$ is included in $%
\zeta _l.$ Nevertheless we will conserve the terms $\sigma _j$ in all the
spectral decompositions.}. Also, as in the pure case: 
\[
(\beta (\overline{\sigma ,\zeta _l}|\beta (\overline{\sigma ^{\prime },\zeta
_{l^{\prime }}}))=\delta _{\sigma \sigma ^{\prime }}\delta _{ll^{\prime
}}\varepsilon _l 
\]
\begin{equation}
(\beta (\widetilde{\sigma ,\zeta _l}|\beta (\widetilde{\sigma ^{\prime }}%
,\zeta _{l^{\prime }}))=\delta _{\sigma \sigma ^{\prime }}\delta
_{ll^{\prime }}\varepsilon _l  \label{4.16}
\end{equation}
where $\varepsilon _l=0$ if $%
\mathop{\rm Im}%
\zeta _l\neq 0$ and $\varepsilon _l=1$ if $%
\mathop{\rm Im}%
\zeta _l=0$, so the states corresponding to complex poles are ghosts as
before. The same thing happens with $\rho (\overline{0,z_n}),$ $\rho (%
\overline{z_n,0}),$ $\rho (\widetilde{0,z_n}),$ and $\rho (\widetilde{z_n,0}%
).$

We can generalize the definition of trace as: 
\begin{equation}
Tr\rho =\sum_i<i|\rho |i>=(\rho |\sum_i|i><i|)=(\rho |\sum_i\beta
(i0))=(\rho |I)  \label{4.18}
\end{equation}
where $\{|i>\}$ is any basis of ${\cal H}$. Now using the inner product (\ref
{4.6}) it can be easily proved that all the trace of all the off diagonal
terms vanish. Therefore the trace of all the ghosts vanish.

Finally equations similar to eqs. (3.1) of paper \cite{Laucast} can be
obtained and using this equation it can be proved that, if $\zeta _l,\zeta
_l^{\prime },\zeta _l^{\prime \prime },...$ are complex, then: 
\begin{equation}
Tr\beta (\overline{\sigma ,\zeta _l})\beta (\overline{\sigma ^{\prime
},\zeta _{l^{\prime }}})=0,\text{ }\beta (\overline{\sigma ,\zeta _l})\beta (%
\overline{\sigma ^{\prime },\zeta _{l^{\prime }}})\beta (\overline{\sigma
^{\prime \prime },\zeta _{l^{\prime \prime }}})....=0  \label{4.17'}
\end{equation}
This equation follows also from eq. (\ref{11.H}) and says that the trace of
the product of two ghosts and that the product of three or more ghosts
vanish.

Since the Liouville space is a Hilbert space ${\cal L=H\oplus (H\otimes H)}$
we will have the Gel'fand triplets: 
\begin{equation}
\Phi _{-}\subset {\cal L}_{-}{\cal \subset }\Phi _{-}^{\times }  \label{4.21}
\end{equation}
\begin{equation}
\Phi _{+}\subset {\cal L}_{+}{\cal \subset }\Phi _{+}^{\times }  \label{4.22}
\end{equation}

Let us observe that, in order to satisfy eq. (\ref{4.B}$_5$) it is
sufficient that the regular part of $\rho _{}\in {\cal S}_{-}\otimes {\cal S}%
_{+}$ (since really in the second factor of eq.(\ref{4.B}) there is a bra
not a ket). Thus, as we will see in more detail in section 12 (cf. eq. (\ref
{11.C}))${\cal S\oplus (}$ $\phi _{-}\otimes \phi _{-})\subset \Phi _{-}$
and ${\cal S\oplus (}\phi _{+}\otimes \phi _{+})\subset \Phi _{+}.$

As we will see $\Phi _{-}$ will be the space of physically admissible
states, precisely the space of states such that they evolve with a
non-decreasing of entropy accordingly to the Second Law of Thermodynamics. $%
\Xi \oplus (\Xi \otimes \Xi )\backslash \Phi _{-}$ is the set of physically
non-admissible states. $\Phi _{+}$ is the space of states such that they
evolve with a non-growing entropy and, therefore they are clearly non
physical. Macroscopically the physically admissible evolution are those that
appear in nature, namely those that begin in an unstable state and go
towards equilibrium (Gibbs ink drop spreading in the glass of water, a sugar
lump solving in a cup of coffee, etc.). We will consider that everything the
same in the microscopical case, namely that $\Phi _{-}$ is the space of
physically realizable states. The physically non admissible evolutions of
space $\Phi _{+}$ can be obtained by the time inversion of the admissible
ones, therefore they begin in an equilibrium state and evolve towards an
unstable state (the ink or the sugar concentrating spontaneously and
creating the drop or the lump). This kind of evolutions does not appear in
nature, because the spontaneous appearance of an unstable state by a
fluctuation, even if no completely impossible (remember we are developing
quantum mechanics, an essentially statistical theory \cite{Ballentine}), is
highly improbable.

For all these reasons we will consider $\Phi _{-}$ the space of physical
states.

\section{Linear operators and observables.}

We will now consider the linear operators $A,$ which are (anti)-linear
functional over $\Phi _{-}$ and therefore belong to $\Phi _{-}^{\times },$
e. g. a derivative operator belongs to this space.

But these linear operators are merely theoretical or mathematical
''observables''. Real physical observables are less subtle, e. g.: there are
not real apparatuses to measure mathematical derivatives. Real physical
devices only measure ratios of small but finite quantities. Therefore we can
consider that real physical observables live in an space endowed at least
with the properties of ${\cal S\oplus (S\otimes S).}$ As in the case of the
state it is also useful that these observable would have some analytic
properties. There are two natural subspaces of ${\cal S\oplus (S\otimes S)}$
with definite analytic properties $\Phi _{-}$ and $\Phi _{+}$. As we will
see, in order to reproduce the relation between the Schroedinger and the
Heisenberg pictures in the new theory, we must choose $\Phi _{+}$ as the
space of regular physical observables$.$ As 
\begin{equation}
\Phi _{-},\Phi _{+}\subset {\cal L}  \label{4.4'}
\end{equation}
the products between vectors of these two spaces are well defined. Then the
mean values of all the observables of $\Phi _{+}$ in the states of $\Phi
_{-} $ are well defined. This property is sufficient to develop quantum
mechanics formalism.

As we will see in a scattering theory, physical states are related to the
preparation, they propagate towards the future in the Schroedinger picture,
and they are well represented by states of space $\Phi _{-\text{, }}$while
physical observables are related with measurements, they propagate towards
the past in the Heisenberg picture, and they are well represented by states
of space $\Phi _{+}$ \cite{AntoniouBohm}.

After all these considerations the mean value of observable $A\in \Phi _{+}$
in the states $\rho \in \Phi _{-}$ is: 
\begin{equation}
<A>_\rho =A[\rho ]=(\rho |A)  \label{4'.5}
\end{equation}
where $A[\rho ]$ is an (anti) linear functional and: 
\[
A=A_0\rho (0)+\sum_n[A_{n0}\rho (\widetilde{z_n,0})+A_{0n}\rho (\widetilde{%
0,z_n})+\int_\Gamma [A_{z0}\rho (\widetilde{z,0})+A_{0z}\rho (\widetilde{0,z}%
)]dz+ 
\]
\begin{equation}
\sum_{jl}A_{jl}\widetilde{\beta (\sigma _j,\zeta _l^j)}+\int_0^\infty
d\sigma [A_\sigma \beta (\sigma )+\sum_lA_{\sigma l}\beta (\widetilde{\sigma
,\zeta _l})+\int_C\rho _{\sigma z}\beta (\widetilde{\sigma ,z})dz]
\label{4'.6}
\end{equation}
where the coefficients $A$ must satisfy eqs. (\ref{4.12'}) and (\ref{4.13}).
To fulfill these conditions, and eqs. (\ref{4.8}) and (\ref{4.12}) for the
coefficients of $\rho ,$ it is sufficient that the analytic continuation of
eq. (\ref{4'.5}) would be possible, one variable in the lower half plane and
the other in the upper half plane, as we have done in ref. \cite{Laucast}.

From the Gel'fand-Maurin theorem \cite{Chiu} we know that we can diagonalize
eq. (\ref{4'.6}) as: 
\begin{equation}
A=\sum_ia_i|a_i><a_i|  \label{4'.7}
\end{equation}
where in general $a_i\in {\Bbb C}{\bf ,}$ $i$ is an index such that the
whole spectrum of $A$ is covered by the sum, and $|a_i>$ belongs to some
specific rigged Hilbert space. If $A$ is selfadjoint then obviously $a_i\in 
{\Bbb R}$ and usually $|a_i>\in {\cal S}^{\times }{\cal \ }$. In particular
we can expand the energy operator as: 
\begin{equation}
H=\sum_ih_i|h_i><h_i|  \label{4'.8}
\end{equation}
where $h_i\in {\Bbb R}$ and $|h_i>\in {\cal S}^{\times }$, that can simply
be obtained from eq. (\ref{2.3}) or eq. (\ref{4.9}) as: 
\begin{equation}
H=\int_0^\infty E|E><E|dE=E_0\beta (0)+\int_0^\infty d\sigma E_\sigma \beta
(\sigma )  \label{4'.9}
\end{equation}
Thus $H$ has two different spectral expansions, both very useful: one as
observable eq. (\ref{4'.8}) and another as a evolution operator eq. (\ref
{3.5}), namely, in the short hand notation of section 3: 
\begin{equation}
H=\sum_iz_i|\overline{i}><\widetilde{i}|  \label{4'.10}
\end{equation}
where $z_i\in {\Bbb C}{\bf .}$ The difference between the two expansions
comes from the fact that really they are weak equation corresponding to: 
\begin{equation}
<\psi _1|H|\psi _2>=\sum_nh_n<\psi _1|h_n><h_n|\psi _2>  \label{4'.11}
\end{equation}
where $|\psi _1>,|\psi _2>\in {\cal S}$, 
\begin{equation}
<\varphi |H|\psi >=\sum_iz_i<\varphi |\overline{i}><\widetilde{i}|\psi >
\label{4'.12}
\end{equation}
where $|\psi >\in \phi _{-}$ and $|\varphi >\in \phi _{+}.$

By now we have all the mathematical objects to formulate our axiomatic
theory.

\section{Axioms.}

We will follow the main lines of ref. \cite{Ballentine}. So we postulate:

{\bf Axiom 1. }To each dynamical variable ${\cal R}$ (physical concept)
there corresponds a linear operator $R$ $\in \Phi _{+\text{ }}\subset {\cal L%
}$ $_{+}$(mathematical object) and the possible values of the dynamical
variable are the eigenvalues of the operator.

{\bf Axiom 2.} To each physical state there corresponds a unique state
operator $\rho \in \Phi _{-}\subset {\cal L}_{-}{\cal .}$ The average value
of the dynamical variable ${\cal R,}$ $($e.g.: of position, momentum,
energy, etc.) represented by the operator $R$, in the virtual ensemble of
events that may result from a preparation procedure for the state,
represented by the operator $\rho ,$ is: 
\begin{equation}
<{\cal R}>_\rho =\frac{Tr[\rho R]}{Tr\rho }  \label{5.1}
\end{equation}

From these axioms, if we postulate the invariance of the theory under
Galilei transformation, the explicit expression of the operators $R$ can be
found and Schroedinger and Liouville equations can be deduced, as in book 
\cite{Ballentine} or in paper \cite{Jordan}. Moreover, Planck's constant $%
\hbar $ appears as a proportionality coefficient between the geometrical
generators and the physical magnitudes. In fact, these deductions can be
implemented since we have just restricted the domain of definition of the
states and of the observables but all the relevant demonstrations of the
quoted reference remain valid. We do not reproduce this demonstration here
because it is not in our main line of reasoning. So, in order to avoid these
demonstrations, even if we maintain the Galilei invariance, we precise the
main features of the time evolution by the following axiom:

{\bf Axiom 3. }The time evolution of a state $\rho (t)\in \Phi _{-}$ is: 
\begin{equation}
\rho (t)=e^{-iHt}\rho (0)e^{iHt}=e^{-iLt}\rho (0)  \label{5.2}
\end{equation}
where $\rho (0),\rho (t)\in \Phi _{-}$ and $H$ is the hamiltonian operator
of the system.

In this equation we must use eq. (\ref{4'.10}) if we want to expand $H.$ The
exponents $iHt$ really read $i\hbar ^{-1}Ht,$ so it is this axiom the one
that introduces the universal constant $\hbar .$ Of course we make $\hbar =1$
below$.$ From this axiom we also can demonstrate that $\rho (t)$ satisfies
Liouville equation (\ref{Liouville}).

Since we have restricted the spaces of definition of the observables and the
states to two spaces which are contained in ${\cal L}$ nothing unphysical
can really happen. Furthermore we obviously retain the main result of the
usual quantum physics, as we will see in the next two sections, but with the
new axiomatic structure we will gain new results that are, in fact,
confirmed by experimental evidences.

\section{First consequences of the axioms.}

We can now obtain the first consequences of the axioms following ref. \cite
{Ballentine}:

a.- As $\rho $ is both in the numerator and in the denominator of eq. (\ref
{5.1}) we can normalize the state as: 
\begin{equation}
Tr\rho =1  \label{6.1}
\end{equation}
b.- If we postulate that projectors like $P=|u><u|$ are observables, and
since their eigenvalues are 0 and 1, we can see that $<P>_\rho =<u|\rho |u>$
is real and positive so (cf. theorem 1 ref. \cite{Ballentine}): 
\begin{equation}
\rho =\rho ^{\dagger },\text{ }<u|\rho |u>\geq 0  \label{6.2}
\end{equation}

c.- As $\rho =\rho ^{\dagger },$ $\rho $ can be expanded as: 
\begin{equation}
\rho =\sum_i\rho _i|\rho _i><\rho _i|  \label{6.3}
\end{equation}
where $\rho _i\in {\Bbb R}${\bf \ }and $|\rho _i>$ belongs to some adequate
rigged Hilbert space. Then from the three first equations of this section we
can obtain that: 
\begin{equation}
\sum_i\rho _i=1,\text{ }\rho _i=\rho _i^{*},\text{ }0\leq \rho _i\leq 1
\label{6.4}
\end{equation}

d.- If we postulate that the mean value of any dynamical variable must be
real, i. e.: 
\begin{equation}
<{\cal R}>_\rho \in {\Bbb R}  \label{6.5}
\end{equation}
then, for any pure state $\rho =|v><v|,$ $|v>\in \phi _{-}$ , we have: 
\begin{equation}
<{\cal R}>_\rho =<v|R|v>\in {\Bbb R}  \label{6.6}
\end{equation}
so, according to theorem 1 of ref.\cite{Ballentine} it is: 
\begin{equation}
R=R^{\dagger }  \label{6.7}
\end{equation}
so we can expand $R$ as in eq. (\ref{4.7}), namely: 
\begin{equation}
R=\sum_nr_n|r_n><r_n|  \label{6.8}
\end{equation}
where $r_n\in {\Bbb R}$ and $|r_n>$ belongs to some adequate rigged Hilbert
space$.$

\section{Probabilities.}

From axiom 1 we have that: 
\begin{equation}
<{\cal R}>_\rho =\sum_nr_np_n(\rho )  \label{7.1}
\end{equation}
where $p_n(\rho )$ is the probability to obtain the measurement $r_n,$ when
we measure the dynamical variable ${\cal R}$ in the quantum state $\rho $.
From axiom 2 we also have: 
\begin{equation}
<{\cal R}>_\rho =R[\rho ]=Tr[(\sum_nr_n|r_n><r_n|)\rho ]=\sum_nr_n<r_n|\rho
|r_n>  \label{7.2}
\end{equation}
In order that the last two equations would be equal it is sufficient that: 
\begin{equation}
p_n(\rho )=<r_n|\rho |r_n>  \label{7.3}
\end{equation}

It can be proved that this condition is also necessary if we repeat the
corresponding demonstration of ref. \cite{Ballentine}.

Then, for every estate $\rho $ and every complete set of commuting
observable $\{R^{(\alpha )}\},$ we can compute the probability to obtain the
measurement $r_n^{(\alpha )}$ for the observable $R^{(\alpha )}$. In fact we
can expand the observable as:

\begin{equation}
R^{(\alpha )}=\sum_nr_n^{(\alpha )}|r_n^{(\alpha )}><r_n^{(\alpha )}|
\label{7.5}
\end{equation}
and the probability is: 
\begin{equation}
p_n^{(\alpha )}(\rho )=<r_n^{(\alpha )}|\rho |r_n^{(\alpha )}>  \label{7.4}
\end{equation}

So we can see that $\rho $ really defines the {\it quantum state }of the
system since, knowing $\rho ,$ we can obtain the probability of any
measurement for any observable of the complete set of commuting observables.
This is, in fact, the maximal information that we can obtain from a quantum
state $\rho $ and, in consequence , this information also defines the
quantum state of the system.

\section{Time-asymmetry and irreversibility.}

In the last two sections we have briefly reviewed some results of ordinary
quantum mechanics that turn out to be also valid in the new theory. It would
be quite boring to continue this road reobtaining well known results so we
will now consider the new features.

We will say that:

{\bf Time-asymmetry} is the property of some single objects that turn out to
be asymmetric, under the action of time-inversion Wigner operator $K,$ e. g.
non real states $|\psi >,$ defined as the states such that $K|\psi >\neq
|\psi >$. In our case these objects are always statistical objects from the
ensemble we are considering, since we are developing a statistical theory.
Therefore the time-asymmetry of particular evolution of the members of the
ensemble will be never taken into account.

{\bf Non time-reversal invariance} is the property of some set of objects
which are not invariant under $K$, e.g. the space $\phi _{-}$ which has the
property (\ref{2.7}).

{\bf Irreversibility} is the property of some physical time evolutions such
that its time inverted evolution turns out to be non-physical, namely it is
physically forbidden \cite{Messiah}, \cite{CGG}.

If we would further precise the term, the just introduced irreversibility
would be the {\bf dynamical irreversibility }(as we will see in a moment
this irreversibility stems directly from the axioms){\bf .} {\bf %
Thermodynamical irreversibility }will be defined as the growing of entropy
in section 13 (and we will see that more elements must be added to define
this notion).

From the just quoted eq. (\ref{2.7}) and the definitions at the beginning of
section 4 we have that: 
\begin{equation}
{\cal K}:\Phi _{-}\rightarrow \Phi _{+}\neq \Phi _{-}  \label{8.1}
\end{equation}
where ${\cal K}\rho =K\rho K^{\dagger }$, and we can see that the physically
admissible quantum states space of the theory is not time-reversal invariant.

From ref. \cite{Laucast}, eq. (4.2) and eq. (4.3), we know that: 
\begin{equation}
e^{-iHt}:\phi _{-}\rightarrow \phi _{-,}\text{ }if\text{ }t>0,\text{ }%
e^{-iHt}:\phi _{+}\rightarrow \phi _{+},\text{ }if\text{ }t<0  \label{8.2}
\end{equation}
Therefore, using the same demonstration regarding now the analytic
properties of the functions of variable $\nu ,$ it can be proved that:

\begin{equation}
e^{-iLt}:\Phi _{-}\rightarrow \Phi _{-,}\text{ }if\text{ }t>0,\text{ }%
e^{-iLt}:\Phi _{+}\rightarrow \Phi _{+},\text{ }if\text{ }t<0  \label{8.3}
\end{equation}
so axiom 3 states that if $\rho (t)\in \Phi _{-}$ its evolution is only
defined for $t>0$, and therefore the evolution operator $e^{-iLt\text{ }}$%
cannot be physically inverted since its mathematical inverted operator $%
e^{iLt}$ corresponds to $t<0$ and therefore it is not well defined within
space $\Phi _{-}.$ Namely the inverted evolution is forbidden by axiom 3.
Therefore we have found that the new theory contains dynamical irreversible
evolutions.

Of course $t=0$ is an arbitrary time so the condition $t>0$ physically
simply means that operators $e^{-iHt\text{ }}$and $e^{-iLt}$ are not well
defined for $t\rightarrow -\infty $ for the state of space $\Phi _{-}$.
Analogously, the condition $t<0$ means that the sane operators are not well
defined for $t\rightarrow +\infty $ for the states of $\Phi _{+}.$

\section{Schr\"{o}dinger and Heisenberg pictures and scattering experiments.}

In the Schroedinger picture, if $\rho (t)$ is the time-variable state of the
system and ${\cal R}$ is a fixed dynamical variable, from axioms 2 and 3 we
have: 
\begin{equation}
<{\cal R}>_\rho (t)=Tr[\rho (t)R]=Tr[e^{-iLt}\rho (0)R]=Tr[e^{-iHt}\rho
(0)e^{iHt}R]  \label{9´.1}
\end{equation}
According to axiom 3 $\rho (t)\in \Phi _{-}$ so, from eq. (\ref{8.3}$_1),$
we know that the last equation is only valid if $t>0$. Now from the cyclic
property of the trace we also have that: 
\begin{equation}
<{\cal R}>_\rho (t)=Tr[\rho (0)e^{iHt}R\text{ }e^{-iHt}]  \label{9'.2}
\end{equation}
So we can define a time-variable Heisenberg operator: 
\begin{equation}
R_H(t)=e^{-iH(-t)}R\text{ }e^{iH(-t)}=e^{-iL(-t)}R  \label{9´.3}
\end{equation}
Then we have the Heisenberg picture equation: 
\begin{equation}
<{\cal R}>_\rho (t)=Tr[\rho (0)R_H(t)]  \label{9'.4}
\end{equation}
But from eq. (\ref{8.3}$_2$) and since $-t<0$ we know that the last time
equation is only valid if $R\in \Phi _{+}.$ This fact justifies both the
choice of the operators space done in axiom 1, and what we have said in
section 4.

In other words: in a scattering experiment (\cite{AntoniouBohm}, \cite
{Diener}) the states are prepared at a time $t_1$ and propagate towards the
future and therefore to times $t>t_1$ so according to eq. (\ref{8.3}$_1$) $%
\rho \in \Phi _{-}.$ At time $t_2>t_1$ dynamical variables ${\cal R}$ are
measured i. e. the S-matrix and the corresponding probabilities are
obtained. But we can invert the procedure and propagate the dynamical
variables ${\cal R}$ and the corresponding operators $R$ towards the past
down to time $t_1<t_2.$ Then we must propagate $R$ towards the past,
therefore according to eq. (\ref{8.3}$_2)$ $R\in \Phi _{+}.$ So now we see
the motivation of the choice of the spaces for $\rho $ and $R$ made in
axioms 1 and 2, namely $\rho \in \Phi _{-}$ and $R\in \Phi _{+}.$

\section{Equilibrium and decoherence.}

We will face the problem of equilibrium in four steps: in the first one we
will obtain a strong limit, in the second one a weak limit, in the third one
the dominant time evolution components, and in the fourth one we will
combine the two last ones to obtain some physical conclusions.

a.-From eqs. (\ref{4.F}) and (\ref{4.15}) we can deduce that if $\rho (t)\in
\Phi _{-}:$%
\[
\rho (t)=\rho _0\rho (0)+\sum_n[\rho _{n0}e^{-i(z_n-\omega _0)t}\rho (%
\overline{z_{n,}0})+\rho _{0n}e^{-i(\omega _0-z_n^{*})t}\rho (\overline{0,z_n%
})+ 
\]
\[
+\int_\Gamma [\rho _{z0}e^{-i(z-\omega _0)t}\rho (\overline{z,0})+\rho
_{0z}e^{-i(\omega _0-z)t}\rho (\overline{0,z})]dz+\sum_{jl}\rho
_{jl}e^{-i\zeta _l^jt}\overline{\beta (\sigma _j,\zeta _l^j)} 
\]
\begin{equation}
\int_0^\infty d\sigma [\rho _\sigma \beta (\sigma )+\sum_l\rho _{\sigma
l}e^{-i\zeta _lt}\beta (\overline{\sigma ,\zeta _l})+\int_C\rho _{\sigma
z}e^{-izt}\beta (\overline{\sigma z})dz]  \label{9.1}
\end{equation}
where $\zeta _l$ and $z_n$ symbolize the complex poles. If we call, as it is
traditional: 
\[
z_n=\overline{\omega }_n-\frac i2\gamma _n,\text{ }\gamma _n>0 
\]
\begin{equation}
\zeta _l=\overline{\nu }_l-i\Gamma _l,\text{ }\Gamma _l>0  \label{9.2}
\end{equation}
we have that: 
\[
\rho (t)=\rho _0\rho (0)+\sum_n[\rho _{n0}e^{-i(\overline{\omega }_n-\omega
_0)t}e^{-\frac 12\gamma _nt}\rho (\overline{z_{n,}0})+\rho _{0n}e^{-i(\omega
_0-\overline{\omega }_n)t}e^{-\frac 12\gamma _nt}\rho (\overline{0,z_n})+ 
\]
\[
+\int_\Gamma [\rho _{z0}e^{-i(z-\omega _0)t}\rho (\overline{z,0})+\rho
_{0z}e^{-i(\omega _0-z)t}\rho (\overline{0,z})]dz+\sum_{jl}\rho _{jl}e^{-i%
\overline{\nu }_l^jt}e^{-\Gamma _l^jt}\overline{\beta (\sigma _j,\zeta _l^j)}%
+ 
\]
\begin{equation}
\int_0^\infty d\sigma [\rho _\sigma \beta (\sigma )+\sum_l\rho _{\sigma
l}e^{-i\overline{\nu }_lt}e^{-\Gamma _lt}\beta (\overline{\sigma ,\zeta _l}%
)+\int_C\rho _{\sigma z}e^{-izt}\beta (\overline{\sigma z})dz]  \label{9.3}
\end{equation}
For complex poles it is $\gamma _n,\Gamma _l>0,$ so terms containing these
positive gammas vanish when $t\rightarrow +\infty .$ The terms corresponding
the continuous spectra do not vanish since the curves $\Gamma $ and $C$ can
be taken to be contained in the real axis almost anywhere. Then we obtain
the strong limit: 
\[
\rho (t)\rightarrow \rho _{*}(t)=\rho _0\rho (0)+\int_\Gamma [\rho
_{z0}e^{-i(z-\omega _0)t}\rho (\overline{z,0})+\rho _{0z}e^{-i(\omega
_0-z)t}\rho (\overline{0,z})]dz+ 
\]
\begin{equation}
\int_0^\infty d\sigma [\rho _\sigma \beta (\sigma )+\int_C\rho _{\sigma
z}e^{-izt}\beta (\overline{\sigma z})dz]  \label{9.4}
\end{equation}
so any state goes to a state of ''dynamical equilibrium''. We use the
adjective ''dynamical'' since it is a final state that it is a function of
time. If we take into account normalization (\ref{6.1}) we have: 
\begin{equation}
Tr\rho (t)=Tr\rho _{*}(t)=1  \label{9.5}
\end{equation}
so: 
\begin{equation}
\rho _0+\int_0^\infty \rho _\sigma d\sigma =1,  \label{9.6}
\end{equation}
This is certainly a equation that the $\rho _0$ and $\rho _\sigma $ must
satisfy, but in principle this is the only condition.

So in general the dynamical equilibrium state is not unique and depends of
the initial conditions. This would be the general case.

b.-Moreover, if we consider that really the $\rho $ are functionals over the
observables $A$ , since only the mean values $<A>_\rho =A[\rho ]$ are
physically observed, and we use the Riemann-Lebesgue theorem, as in ref. 
\cite{LaucastIII}, we obtain in a weak sense that: 
\begin{equation}
\rho (t)\rightarrow \rho _{*}=\rho _0\rho (0)+\int_0^\infty \rho _\sigma
\rho (\sigma )d\sigma  \label{9.6'}
\end{equation}
Therefore only the terms on the diagonal remain and we obtain a stationary
final equilibrium as a weak limit. Only the state $\rho (0),$ corresponding
to the ground state of the discrete spectrum (e. g. an oscillator), and the
states $\rho (\sigma )$, corresponding to the diagonal states of the
continuous one (e. g. the bath), remain in equilibrium. Thus we have proved
that quantum decoherence takes place in our theory.

c.- But, using this method, we just obtain the limit, but we cannot see how
this limit is attained. So we will use another approach. We know that the
dominant component of the evolution towards equilibrium is given by the pole
terms \cite{Sudarshan}. In fact, this component is an excellent
approximation for intermediate times: not too short times, so the Zeno
effect would not be important, not too long times, so the Khalfin effect
would not be important. Furthermore, experimentally we know that this
intermediate period turns out to be very long, since, up to now, the Khalfin
effect is not detected. So if we want to have a very good approximation of
the evolution towards equilibrium we can neglect the regular background
fields terms of curves $\Gamma $ and $C$ and only consider the poles terms
and the singular diagonal terms, precisely:

\[
\rho (t)=\rho _0\rho (0)+\sum_n[\rho _{n0}e^{-i(\overline{\omega }_n-\omega
_0)t}e^{-\frac 12\gamma _nt}\rho (\overline{z_{n,}0})+\rho _{0n}e^{-i(\omega
_0-\overline{\omega }_n)t}e^{-\frac 12\gamma _nt}\rho (\overline{0,z_n}) 
\]
\begin{equation}
\sum_{jl}\rho _{jl}e^{-i\overline{\nu }_l^jt}e^{-\Gamma _l^jt}\overline{%
\beta (\sigma _j,\zeta _l^j)}+\int_0^\infty d\sigma [\rho _\sigma \beta
(\sigma )+\sum_l\rho _{\sigma l}e^{-i\overline{\nu }_lt}e^{-\Gamma _lt}\beta
(\overline{\sigma ,\zeta _l})]  \label{10.9}
\end{equation}
We will call: 
\begin{equation}
\rho _{*}=\rho _0\rho (0)+\int_0^\infty \rho _\sigma \rho (\sigma )d\sigma
\label{10.11}
\end{equation}
\[
\text{ }e^{-\gamma _{\min }t}\rho _1(t)=\sum_n[\rho _{n0}e^{-i(\overline{%
\omega }_n-\omega _0)t}e^{-\frac i2\gamma _nt}\rho (\overline{z_{n,}0})+\rho
_{0n}e^{-i(\omega _0-\overline{\omega }_n)t}e^{-\frac i2\gamma _nt}\rho (%
\overline{0,z_n})+ 
\]
\begin{equation}
\sum_{jl}\rho _{jl}e^{-i\overline{\nu }_l^jt}e^{-\Gamma _l^jt}\overline{%
\beta (\sigma _j,\zeta _l^j)}+\int_0^\infty d\sigma \sum_l\rho _{\sigma
l}e^{-i\overline{\nu }_lt}e^{-i\Gamma _l}\beta (\overline{\sigma ,\zeta _l})]
\label{10.11'}
\end{equation}
where we have made explicit the minimum of the $\gamma $ and the $\Gamma $
in the l.h.s. of the second equation$.$ Since $\gamma ,\Gamma >0$ , when $%
t\rightarrow \infty $ we have: 
\begin{equation}
\rho (t)\rightarrow \rho _{*}  \label{10.12}
\end{equation}
So, in this case, we obtain the usual stationary equilibrium which is not
time dependent (but it still depends on the initial condition, through the $%
\rho _0$ and the $\rho _\sigma ,$ we will find an equilibrium independent of
these conditions in section 14). The normalization condition is still eq. (%
\ref{9.6}) and we have: 
\begin{equation}
Tr\rho _1(t)=0  \label{10.13}
\end{equation}
which is also a consequence of the fact that the ghost has vanishing trace.

So again we have obtained the usual equilibrium state and, as the
off-diagonal terms vanish when $t\rightarrow \infty ,$ the phenomenon of
decoherence is also proved.

The present method has been used to study decoherence in the cosmological
case in papers \cite{Lombardo}.

\section{Conservation of the norm, the trace, and the energy. Lyapunov
variables. Survival probability.}

a.- In eq. (\ref{9.3}) we have shown that some states of the theory vanish
for $t\rightarrow +\infty $, precisely the ''ghost'' states such that $%
\gamma _n,\Gamma _l>0.$ Then we could wonder if the trace, the norm, and the
energy are conserved in our theory. In fact, it is so, since we know that
the trace of the off diagonal terms vanishes, so we have: 
\begin{equation}
Tr\rho (t)=\rho _0+\int_0^\infty \rho _\sigma d\sigma =1  \label{11.1}
\end{equation}
Also: 
\begin{equation}
Tr[\rho (t)H]=\omega _0\rho _0+\int_0^\infty \sigma \rho _\sigma d\sigma
=const  \label{11.2}
\end{equation}
so the trace and the mean value of the energy are constant

In the pure state case these equations read: 
\begin{equation}
<\psi |\psi >=1  \label{11.3}
\end{equation}
\begin{equation}
<\psi |H|\psi >=const.  \label{11.4}
\end{equation}
so the norm of pure states $|\psi >$ are also a constant.

b.- To continue let us repeat the reasonings of the beginning of this
section in a different case : we will use another basis (precisely the one
that can be obtained by the products of the basis of section 3) and the
shorthand notation of section 3 and let it be $N\neq 0$. Let us consider the
space ${\cal S\oplus (}\phi _{-}\otimes \phi _{-\text{ }})$ and a state $%
\rho \in $ ${\cal S\oplus (}\phi _{-}\otimes \phi _{-\text{ }})$ , that we
can develop as: 
\begin{equation}
\rho =\sum_i\rho _i|i><i|+\sum_{ij}\rho _{ij}|\overline{i}><\overline{j}|
\label{11.A}
\end{equation}
where $\rho _i\in {\cal S,}$ $\rho _{ij}\in {\cal S}_{-}\otimes {\cal S}_{+}$%
. Now as: 
\begin{equation}
\rho _{ij}=\rho _{\sigma +\frac 12\nu ,\sigma -\frac 12\nu }=\rho _{\rho \nu
}  \label{11.B}
\end{equation}
it turns out that $\rho \in \Phi _{-},$ since $\rho _{\sigma \nu }$ as a
function of $\nu $ belongs to ${\cal S}_{-}^{(\sigma )},$ then we can
conclude that: 
\begin{equation}
{\cal S\oplus (}\phi _{-}\otimes \phi _{-\text{ }})\subset \Phi _{-}\subset 
{\cal L}_{-}{\cal \subset }\Phi _{-}^{\times }\subset {\cal S}^{\times
}\oplus (\phi _{-}^{\times }\otimes \phi _{-}^{\times })  \label{11.C}
\end{equation}
So any function $\beta \in \Phi _{-}^{\times }$ (let say $\rho \in \Phi _{-}$
or $\beta \overline{(\sigma ,\zeta _l)}\in \Phi _{-}^{\times })$ can also be
expanded as in eq. (\ref{11.A}) and then: 
\begin{equation}
\beta (t)=e^{-iHt}(\sum_i\beta _i|i><i|+\sum_{ij}\beta _{ij}|\overline{i}><%
\overline{j}|)e^{iHt}=\sum_i\beta _i|i><i|+\sum_{ij}\beta
_{ij}e^{-i(z_i-z_j^{\star })t}|\overline{i}><\overline{j}|  \label{11.D}
\end{equation}
Then as usual, if we call: 
\begin{equation}
z_i=\omega _i-\frac i2\gamma _i,\text{ }\gamma _i>0  \label{11.E}
\end{equation}
we have: 
\[
\beta (t)=\sum_i\beta _i|i><i|+\sum_{ij}\beta _{ij}e^{-i(\omega _j-\omega
_i)t}e^{-\frac 12(\gamma _i+\gamma _j)}|\overline{i}><\overline{j}|= 
\]
\begin{equation}
\sum_i\beta _i|i><i|+\sum_{IJ}\beta _{IJ}e^{-i(\omega _J-\omega _I)t}|\omega
_I><\omega _J|+\sum_{ij\neq IJ}\beta _{ij}e^{-i(\omega _j-\omega _i)t}e^{-%
\frac 12(\gamma _i+\gamma _j)t}|\overline{i}><\overline{j}|  \label{11.F}
\end{equation}
where again the indices $I,J,...$ correspond to the real poles of the real
continuous spectrum ($ij\neq IJ$ means that either $i$ or $j$ or both
correspond to ''ghost'' states) and as $\gamma _i+\gamma _j>0,$ so if we
make $t\rightarrow \infty $ we can see that we can expand $\beta _{*}(t)$
and $e^{-\gamma _{\min }t}\beta _1(t)$ as: 
\begin{equation}
\beta _{*}(t)=\sum_i\beta _i|i><i|+\sum_{IJ}\beta _{IJ}e^{-i(\omega
_J-\omega _I)t}|\omega _I><\omega _J|  \label{11.G}
\end{equation}
\begin{equation}
e^{-\gamma _{\min }t}\beta _1(t)=\sum_{ij\neq IJ}\beta _{ij}e^{-i(\omega
_j-\omega _i)t}e^{-\frac 12(\gamma _i+\gamma _j)t}|\overline{i}><\overline{j}%
|  \label{11.H}
\end{equation}
Finally on this basis the conservation of the norm and the energy read: 
\begin{equation}
Tr\beta (t)=\sum_i\beta _i=1  \label{11.H'}
\end{equation}
\begin{equation}
Tr(\beta (t)H)=\sum_i\omega _i\beta _i=const.  \label{11.H''}
\end{equation}
so we have proved, in another basis, that the trace and the mean value of
the energy are constant.

So up to now, every scalar we have introduced is time-constant and it seems
impossible to define Lyapunov variables. But again let us try to find these
results, using now another approach: considering only the regular component $%
\rho _{reg}\in {\cal H\otimes H}$ of $\rho $ and using the restriction of
the trace on the regular component, namely: 
\begin{equation}
tr\rho _{reg}=\sum_i<i|\rho _{reg}|i>=\sum_i<\widetilde{i}|\rho _{reg}|%
\overline{i}>  \label{11.H'''}
\end{equation}
where the last equation can be obtained by analytic continuation. Then we
have: 
\begin{equation}
tr\rho _{reg}(t)=\sum_i\rho _{ii}e^{-\gamma _it}<\overline{i}|\overline{i}%
>=\sum_i\rho _{ii}e^{-i\gamma _it}\varepsilon _i=\sum_I\rho _{II}=const.
\label{11.J}
\end{equation}
\begin{equation}
tr[\rho _{reg}(t)H]=\sum_i\rho _{ii}(\overline{\omega }_i-\frac i2\gamma
_i)e^{-\gamma _it}<\overline{i}|\overline{i}>=\sum_i\rho _{ii}(\overline{%
\omega }_i-\frac i2\gamma _i)e^{-i\gamma _it}\varepsilon _i=\sum_I\omega
_I\rho _{II}=const.  \label{11.K}
\end{equation}
so again everything is constant and we cannot find Lyapunov variables.

From eq. (\ref{11.H}) we can now prove eq. (\ref{4.17'}). In fact, in $\rho
_1(t)$ or $\beta _1(t)$ there are only terms like $|no-ghost><\overline{ghost%
}|,$ $|\overline{ghost}><no-ghost|,$ and $|\overline{ghost}><\overline{ghost}%
|.$ If we compute $\rho _1^2(t)$ or $\beta _1^2(t)$ only the combination $|%
\overline{ghost}><no-ghost|$ $no-ghost><\overline{ghost}|$ survives, so
these squares have only $|\overline{ghost}><\overline{ghost}|$ terms and
their traces vanish. Similarly we can prove that the powers higher than two
of $\rho _1(t)$ or $\beta _1(t)$ vanish.

c.- We have just learnt that the usual constants of quantum mechanics remain
constant in the new theory so they are not Lyapunov variables, namely
variable never-decreasing quantities. Nevertheless we can define Lyapunov
variables in the new theory if we introduce the new ''metric'' operators: 
\begin{equation}
\widetilde{M}=\sum_i|\widetilde{i}><\widetilde{i}|,\text{ }\overline{M}%
=\sum_i|\overline{i}><\overline{i}|  \label{11.5}
\end{equation}
The role of these operators is to exchange bases $\{|\widetilde{i}>\}$ and $%
\{|\overline{i}>\}$ since from eq. (\ref{3.6}) we have: 
\begin{equation}
\widetilde{M}|\overline{i}>=\sum_j|\widetilde{j}><\widetilde{j}|\overline{i}%
>=|\widetilde{i}>  \label{11.6}
\end{equation}
\begin{equation}
\overline{M}|\widetilde{i}>=\sum_j|\overline{j}><\overline{j}|\widetilde{i}%
>=|\overline{i}>  \label{11.7}
\end{equation}
this would happen, e. g. in the $\beta (t)$ of eq. (\ref{11.D}). Therefore: 
\begin{equation}
\widetilde{M}:\phi _{-}\rightarrow \phi _{+},\text{ }\overline{M}:\phi
_{+}\rightarrow \phi _{-}  \label{11.8}
\end{equation}
Then from eq. (\ref{3.6}) we have: 
\begin{equation}
\overline{M}\widetilde{M}=\widetilde{M}\overline{M}=1  \label{11.8'}
\end{equation}
\begin{equation}
<\widetilde{i}|\overline{M}|\widetilde{j}>=\delta _{ij}  \label{11.9}
\end{equation}
\begin{equation}
<\overline{i}|\widetilde{M}|\overline{j}>=\delta _{ij}  \label{11.10}
\end{equation}
so the role of operators $M$ and $\widetilde{M}$ is to eliminate the
''ghost'' factor $\varepsilon _i$ from eqs. (\ref{3.8}), (\ref{3.9}) and, as
we can see, to make a variable what was previously a constant.

Then from eq. (\ref{9.3}) we now have:

\begin{equation}
tr[\rho _{reg}(t)\widetilde{M}]=\sum_i\rho _{ii}e^{-\gamma _it}<\overline{i}|%
\widetilde{M}|\overline{i}>=\sum_I\rho _{II}+\sum_{i\neq I}\rho
_{ii}e^{-\gamma _it}=  \label{11.11}
\end{equation}
\[
const.+\sum_{i\neq I}\rho _{ii}e^{-\gamma _it}=-Y(t) 
\]
where now $Y(t)$ is a Lyapunov variable, if e. g.: $\rho _{ii}\geq 0$ , and
we have: 
\begin{equation}
\stackrel{\bullet }{Y}(t)=\sum_{i\neq I}\rho _{ii}\gamma _ie^{-\gamma _it}>0
\label{11.12}
\end{equation}
because from eq.(\ref{9.2}) $\gamma _i>0$ if $i\neq I.$

If we want to find a Lyapunov variable without the condition $\rho _{ii}\geq
0$ we can introduce the ''linear entropy'' \cite{Gell'man}: 
\begin{equation}
tr[(\rho _{reg}(t)\widetilde{M})^{\dagger }\rho _{reg}(t)\widetilde{M}%
]=\sum_I|\rho _{II}|^2+\sum_{i\neq I}|\rho _{ii}|^2e^{-\gamma _it}=-Y_G(t)
\label{11.11'}
\end{equation}
and in this case we surely have $Y_G^{\bullet }(t)>0$ since $|\rho
_{ii}|^2>0.$

d.- We can also compute survival probabilities and found that in the theory
we have non regular states with pure exponential survival probability. In
fact let: 
\begin{equation}
\rho (t)=|\overline{i(t)}><\overline{i(t)}|  \label{11.13}
\end{equation}
$|\overline{i(t)}>\in \phi _{+}^{\times }.$ At $t=0$ we can consider the
observable $R=|\psi ><\psi |\in \Phi _{+}$ and see how the probability $%
p_i(t),$ to measure the state $\rho (t)$ in the eigenstate $|\psi ),$
evolves: Using eq. (\ref{7.3}) we found: 
\begin{equation}
p_i(t)=<\psi |\overline{i(t)}><\overline{i(t)}|\psi >=e^{-\gamma _it}<\psi |%
\overline{i(0)}><\overline{i(0)}|\psi >  \label{11.14}
\end{equation}
which is an exponentially decaying survival probability with mean life $%
\gamma _i^{-1}.$ This life-time turns out to be infinite for the non-ghost
states $|\omega _I>,$ since in this case $\gamma _I=0.$ Then the non-ghost
states are stable states and the ghost are unstable decaying states. The
physical meaning of the exponents $\gamma _i$ is now clear, they are the
inverse of the mean-life of the decaying processes within the theory.

In the case of the generic state of eq. (\ref{11.F}) and a generic operator $%
R=\sum_\alpha \psi _\alpha |\psi _\alpha ><\psi _\alpha |$ we have:

\[
p_\rho (t)=\sum_{ij\alpha }\rho _{ij}\psi _\alpha e^{-i(\omega _j-\omega
_i)}e^{-\frac 12(\gamma _i+\gamma _j)t}<\psi _\alpha |\overline{i}><%
\overline{j}|\psi _\alpha >= 
\]
\[
\sum_{IJ\alpha }\rho _{IJ}\psi _\alpha <\psi _\alpha |I><J|\psi _\alpha
>e^{-i(\omega _J-\omega _I)t}+ 
\]

.

\begin{equation}
\sum_{ij\neq I}\rho _{ij}\psi _\alpha e^{-i(\omega _j-\omega _i)t}e^{-\frac 1%
2(\gamma _i+\gamma _j)t}<\psi _\alpha |\overline{i}><\overline{j}|\psi
_\alpha >  \label{11.15}
\end{equation}
The second term of the r.h.s. is clearly the dominant exponential component
of the survival probability while the first one gives rise to the Zeno and
Khalfin effects, as can be see in the examples \cite{Arbo}.

\section{Entropy.}

As we have seen the irreversibility of the time evolution of the new theory
allows us to introduce Lyapunov variables. In particular the equilibrium
state $\rho _{*}(t)$ defined in section 11 can be used to define a very
important Lyapunov variable: a quantum conditional entropy \cite{Mackey}.
This entropy coincides with the usual conditional entropy in the classical
limit, which has a remarkable property: it never decreases under a generic
evolution driven by a Markov operator. It is, therefore, an excellent
candidate for the phenomenological internal entropy. In this paper quantum
conditional entropy is just the quantum analogue of classical conditional
entropy, a extensive ever-growing Lyapunov variable that vanishes at
equilibrium. A rigorous and complete study that relates this entropy with
other kinds of entropy is missing nowadays (anyhow see \cite{CGG}, \cite
{Leibowitz}, \cite{Polemica}).

The naive definition of quantum conditional entropy of the state $\rho $
would be: 
\begin{equation}
S=-Tr[\rho \log (\rho \rho _{*}^{-1})]  \label{12.1}
\end{equation}
where $\rho _{*}$ is a equilibrium state like the one of eq. (\ref{10.11}).
Of course, at equilibrium we have $S=0$.

In this section we will use the time evolution based in the dominant pole
component of eq. (\ref{10.9}) (so $N=0)$:

\begin{equation}
\rho (t)=\rho _0\rho (0)+\sum_n[\rho _{n0}e^{-i(\overline{\omega }_n-\omega
_0)t}e^{-\frac 12\gamma _nt}\rho (\overline{z_{n,}0})+\rho _{0n}e^{-i(\omega
_0-\overline{\omega }_n)t}e^{-\frac 12\gamma _nt}\rho (\overline{0,z_n})+
\label{12.2}
\end{equation}
\[
\sum_{jl}\rho _{jl}e^{-i\overline{\nu }_l^jt}e^{-\Gamma _l^jt}\overline{%
\beta (\sigma _j,\zeta _l^j)}+\int_0^\infty d\sigma [\rho _\sigma \beta
(\sigma )+\sum_l\rho _{\sigma l}e^{-i\overline{\nu }_lt}e^{-\Gamma _lt}\beta
(\overline{\sigma ,\zeta _l})]=\rho _{*}+e^{-\gamma _{\min }t}\rho _1(t) 
\]
where, in the second term of the r. h. s. we have made explicit the slowest
dumping factor so $\gamma _{\min }=\min (\gamma _n,\Gamma _l)>0$. All other
factors contained in eq. (\ref{12.2}) are oscillatory and they have a faster
decrease than the slowest dumping factor. Since the trace of the off
diagonal vanishes and using eq. (\ref{4.17'}), we know that: 
\begin{equation}
Tr\rho _1(t)=0,\text{ }Tr\rho _1^2(t)=0,\text{ }\rho _1^n(t)=0,\text{ }if%
\text{ }n>2  \label{12.3}
\end{equation}
We can also introduce a diagonal factor $\rho _{*}^{-1}$ among the factors $%
\rho _1(t)$ and the result will be the same, since being these matrices $%
\rho _{*}^{-1}$ diagonal they only modify the coefficient in the expansion
of $\rho _1(t)$ while eq. (\ref{4.17'}) remains valid. Introducing eq. (\ref
{12.2}) in eq. (\ref{12.1}) we have: 
\begin{equation}
S(t)=-Tr[(\rho _{*}(t)+e^{-\gamma _{\min }t}\rho _1(t))\log (1+e^{-\gamma
_{\min }t}\rho _1(t)\rho _{*}^{-1}(t))]  \label{12.4}
\end{equation}
and expanding the logarithm we obtain: 
\begin{equation}
S(t)=-\frac 12e^{-2\gamma _{\min }t}Tr[\rho _1^2(t)\rho _{*}^{-1}(t)]+...
\label{12.5}
\end{equation}
where the dots symbolize terms with higher powers of $\rho _1(t).$ Then from
eq. (\ref{12.3}) we can conclude that $S(t)=0$, namely that this naive
entropy is constant and it always coincides with equilibrium entropy. This
result is analogous to those obtained in eq. (\ref{11.1}) or eq. (\ref{11.2}%
). So we can conclude, as in the pervious section, that if we do not use the
metric operators $\overline{M}$ and $\widetilde{M}$, or eventually some
projector $P,$ we will always obtain constant naive quantities. Thus we must
introduce, somehow, these operators in the naive definition (\ref{12.1}).
This is the new element that must be added to obtain a growing entropy as
announced in section 9.

The most general and satisfactory solution is to define a projector: 
\begin{equation}
P:\Phi _{-}\rightarrow \Phi _P\subset {\cal L}  \label{A1}
\end{equation}
and consider new projected density matrices $\widetilde{\rho }=P\rho $ in
such a way that a set of necessary properties should be fulfilled. These
properties are:

1.- 
\begin{equation}
Tr\widetilde{\rho }(t)=Tr\widetilde{\rho }_{*}=1\Longrightarrow Tr\widetilde{%
\rho }_1(t)=0  \label{A.2}
\end{equation}
in order that the new projected matrices $\widetilde{\rho }$ would have the
same normalization properties as the old $\rho .$ In short: $Tr\rho =Tr%
\widetilde{\rho }.$ In this sense $P$ is completely different than $%
\overline{M}$ or $\widetilde{M}$ that transform a constant trace into a
variable one.

2.- 
\begin{equation}
Tr\widetilde{\rho }_1^2\neq 0,\text{ }\widetilde{\rho }_1^n(t)\neq 0,n>2
\label{A.3}
\end{equation}
and this must also be the case if some factors $\widetilde{\rho }_{*}^{-1}$
would be included in the product. From these properties we would obtain the
growing of the entropy.

3.- 
\begin{equation}
\widetilde{\rho }_{*}=P\rho _{*}=\rho _{*}  \label{A.4}
\end{equation}
in such a way that in the limit $t\rightarrow \infty $ the projection $P$
turns out to be irrelevant and we reobtain the usual Gibbs entropy for the
equilibrium.

The most general $P$ acting on $\Phi _{-}$ would be: 
\begin{equation}
P=\sum_{ij}p_{ij}|\overline{i})(\widetilde{j}|+\sum_{ij}q_{ij}|\widetilde{i}%
)(\widetilde{j|}  \label{A.5}
\end{equation}
where, for the sake of simplicity we have, once more, trivialized the
notation and we have written the basis $\{\beta (\overline{..})\}$ and $%
\{\beta (\widetilde{..})\}$ with just one index as $\{|\overline{i})\}$ and $%
\{|\widetilde{i})\}$. The most general matrix of $\Phi _{-}$ reads: 
\begin{equation}
\rho =\sum_i\rho _i|\overline{i})  \label{A.6}
\end{equation}
namely eq. (\ref{4.9}) in the new notation. If $p_{ij},q_{ij}\in {\cal %
S\otimes S\otimes S\otimes S}$ then the behavior at infinities is good
enough and then $P\rho \in \Phi _P\subset {\cal L}.$ Now if we symbolize the
diagonal states by $|I)$ and the off diagonal ghost by $|\overline{\mu }) $
we have that: 
\begin{equation}
\rho =\sum_I\rho _I|I)+\sum_\mu \rho _\mu |\overline{\mu })  \label{A.7}
\end{equation}
and 
\[
P=\sum_{IJ}p_{IJ}|I)(J|+\sum_{I\mu }p_{I\mu }|I)(\widetilde{\mu }|+\sum_{\mu
I}p_{\mu I}|\overline{\mu })(I|+ 
\]
\begin{equation}
\sum_{\mu \nu }p_{\mu \nu }|\overline{\mu })(\widetilde{\nu }|+\sum_{\mu
I}q_{\mu I}|\widetilde{\mu })(I|+\sum_{\mu \nu }q_{\mu \nu }|\widetilde{\mu }%
)(\widetilde{\nu }|  \label{A.8}
\end{equation}

But in order to satisfy property 3 it must be: 
\begin{equation}
p_{IJ}=\delta _{IJ},\text{ }p_{\mu I}=q_{\mu I}=0  \label{A.9}
\end{equation}
Let us now compute: 
\begin{equation}
P\rho =\sum_I\rho _I|I)+\sum_{I\mu }p_{I\mu }\rho _\mu |I)+\sum_{\mu \nu
}p_{\mu \nu }\rho _\nu |\overline{\mu })+\sum_{\mu \nu }q_{\mu \nu }\rho
_\nu |\widetilde{\mu })  \label{A.10}
\end{equation}
and let us compute the traces of $\rho $ and $P\rho :$%
\begin{equation}
Tr\rho =\sum_I\rho _I,\text{ }TrP\rho =\sum_I\rho _I+\sum_{I\mu }p_{I\mu
}\rho _\mu  \label{A.11}
\end{equation}
Accordingly to condition 1 these norms must be equal for every $\rho _\mu $
thus: $p_{I\mu }=0.$

Also, as $P^2=P$, it must be: 
\begin{equation}
\text{ }\sum_\lambda p_{\mu \lambda }p_{\lambda \nu }=p_{\mu \nu }\text{ }%
\sum_\lambda q_{\mu \lambda }p_{\lambda \nu }=q_{\mu \nu }  \label{A.12}
\end{equation}
Then these are the conditions that the coefficients of the projector must
satisfy. These equations have solutions, since the first one is satisfied
for any projector in the ''ghost'' space and the second would be satisfied
if $p_{\mu \nu }=q_{\mu \nu }.$ But, of course, more general solutions can
be found. So the solution is not unique and we would have many possible
projectors that will originate many possible entropies (out of equilibrium),
as we will see. We also must remark that the condition $P^2=P$ is not
strictly necessary, so we also can develop a theory where $P$ is not a
projector. Finally: 
\begin{equation}
P\rho _1=\sum_{\mu \nu }p_{\mu \nu }\rho _\nu |\overline{\mu })+\sum_{\mu
\nu }q_{\mu \nu }\rho _\nu |\widetilde{\mu })  \label{A.13}
\end{equation}
Thus $Tr\widetilde{\rho }_1^2=Tr(P\rho _1)^2$ $\neq 0,$ $\widetilde{\rho }%
_1^n(t)\neq 0,n>2$ and these results will be valid if we intercalate $\rho
_{*}^{-1}$ factors. In fact, the ''q'' term in $P$ introduces the $|%
\widetilde{ghost}>$ that does not vanish when multiplied by the $|\overline{%
ghost}>.$ Then condition 2 is also satisfied.

Now we can redefine our quantum conditional entropy as: 
\begin{equation}
S=-Tr[\widetilde{\rho }\log (\widetilde{\rho }\rho _{*}^{-1})]  \label{12.10}
\end{equation}

Repeating the calculations we obtain: 
\begin{equation}
S(t)=-\frac 12e^{-2\gamma _{\min }t}Tr[(\widetilde{\rho }_1(t))^2\rho
_{*}^{-1}(t)]+...  \label{12.11}
\end{equation}
where now the dots symbolize terms that vanish faster than the first one.
But with this new definition $S(t)\neq 0$ and: 
\begin{equation}
\lim_{t\rightarrow \infty }S(t)=0  \label{12.12}
\end{equation}
namely $S(t)$ goes to the equilibrium entropy when $t\rightarrow \infty .$
So, from eq. (\ref{12.11}) we can say that near to equilibrium our quantum
conditional entropy is negative and it always grows, but of course, far from
the equilibrium, the evolution may have not these properties.

So we realize that the ever-growing property of the entropy is not a quantum
property but just a classical one. Thus, for the sake of completeness we
will sketch in the appendix the relation between the classical case and the
quantum case. Then, using our equations we will see that the classical
analogue of our quantum conditional entropy never decreases.

But, for each $P$ there is a different entropy, both in the quantum and the
classical cases. Really this is not a problem since all these entropies
coincide in the equilibrium limit, due to condition 3 (also near equilibrium
all of them have the same dominant dumping factor). So we have one and only
one equilibrium thermodynamic entropy, the Gibbs one, and many non
equilibrium entropies, that depend on the choice of $P,$ namely the choice
of the apparatus that measures these entropies. So our position is that,
even if the arrow of time is intrinsic and defined by space $\Phi _{-}$ the
definition of the out-of-equilibrium entropy is observer (or measurement
apparatus) dependent. But all these entropies grow in the same time
direction and therefore share the same arrow of time.

\section{Thermalization}

From eq. (\ref{3.2}) we know that our method allows to split the hamiltonian
in two non interacting parts a discrete one related with the oscillating and
damped modes of the eventual of $n$ oscillators of our model and a
continuous one , related with the field or the bath $\omega .$ In eq. (\ref
{3.2}) there is no interaction between the discrete and continuous modes so
we may ask ourselves how a bath in thermal equilibrium can thermalize the
oscillators in our model.

Obviously the thermalization can only be seen if we go to the basis where
there is some interaction between the oscillators and the bath, namely the
basis of the unperturbed hamiltonian $H_0$ that we can diagonalize as: 
\begin{equation}
H_0=\sum_nE_n^{(0)}|E_n^{(0)}><E_n^{(0)}|+\int_0^\infty
E^{(0)}|E^{(0)}><E^{(0)}|dE^{(0)}  \label{T.1}
\end{equation}
while the perturbed hamiltonian reads: 
\begin{equation}
H=\sum_nz_n|\overline{f_n}><\widetilde{f_n}|+\int_0^\infty \omega |\overline{%
f_\omega }><\widetilde{f_\omega }|d\omega  \label{T.2}
\end{equation}
But, $H=H_0+V$ so, in the basis $\{|E_n^{(0)},E^{(0)}>\},$ we can see the
interaction and the thermalization phenomena, but not in the basis of eq. (%
\ref{T.2})

In fact, any initial $\rho ,$ in a thermal bath, can be written as: 
\begin{equation}
\rho =\sum_{nm}\rho _{nm}|E_n^{(0)}><E_m^{(0)}|+Z\int_0^\infty e^{-\beta
E^{(0)}}|E^{(0)}><E^{(0)}|dE^{(0)}  \label{T.3}
\end{equation}
(in this section we do not consider the off-diagonal components of the bath,
since we consider that the bath is always in equilibrium, so these
components take no part in the thermalization procedure). The unperturbed
bath is a thermal state at temperature $\beta ^{-1}$ and $Z$ is a
normalization coefficient. Then, introducing: 
\begin{equation}
I=\sum_{n^{\prime }}|\overline{f_{n^{\prime }}}><\widetilde{f_{n^{\prime }}}%
|+\int_0^\infty |\overline{f_\omega }><\widetilde{f_\omega }|d\omega
\label{T.4}
\end{equation}
in each term and tracing away the bath, because we are only interested in
the oscillator's modes term $\rho _{O\text{ ,}}$we have. 
\[
\rho _O=\sum_{nm}\rho _{nm}\sum_{n^{\prime }m^{\prime }}|\overline{%
f_{n^{\prime }}}><\widetilde{f_{n^{\prime }}}|E_n^{(0)}><E_m^{(0)}|\overline{%
f_{m^{\prime }}}><\widetilde{f_{m^{\prime }}|}+ 
\]
\[
Z\int_0^\infty e^{-\beta E^{(0)}}\sum_{n^{\prime }m^{\prime }}|\overline{%
f_{n^{\prime }}}><\widetilde{f_{n^{\prime }}}|E^{(0)}><E^{(0)}|\overline{%
f_{m^{\prime }}}><\widetilde{f_{m^{\prime }}|}dE^{(0)}= 
\]
\[
\sum_{n^{\prime }m^{\prime }}|\overline{f_{n^{\prime }}}><\widetilde{%
f_{m^{\prime }}}|[\sum_{nm}\rho _{nm}<\overline{f_{n^{\prime }}}%
|E_n^{(0)}><E_m^{(0)}|\widetilde{f_{m^{\prime }}}>+ 
\]
\begin{equation}
Z\int_0^\infty e^{-\beta E^{(0)}}<\overline{f_{n^{\prime }}}%
|E^{(0)}><E^{(0)}|\widetilde{f_{m^{\prime }}}>dE^{(0)}]  \label{T.5}
\end{equation}
To see what really is going on let us go to the Friedrichs model of ref. 
\cite{Laucast} and consider the case where $V\rightarrow 0$ and $%
t\rightarrow \infty ,$ using the equations (5.20) of this paper, which
translated to our notation reads: 
\[
|<\overline{f_{n^{\prime }}}|E_n^{(0)}>|^2\rightarrow 0,\text{ }|<E_m^{(0)}|%
\widetilde{f_{m^{\prime }}}>|^2\rightarrow 0 
\]
\begin{equation}
|<\overline{f_{n^{\prime }}}|E^{(0)}>|^2\rightarrow \sim \delta
(E^{(0)}-E_{n^{\prime }}^{(0)}),\text{ }|<E^{(0)}|\widetilde{f_{m^{\prime }}>%
}|^2\rightarrow \sim \delta (E^{(0)}-E_{m^{\prime }}^{(0)})  \label{T.6}
\end{equation}
where we have neglected the terms like $|<\overline{f_{n^{\prime }}}%
|E^{(0)}>|^2$ since they vanish in the limit when $V\rightarrow 0$ because
they relate the discrete eigenkets with the continuous ones. Using these
equations we obtain the result: 
\begin{equation}
\rho _0\rightarrow \sim \sum_n|\overline{f_n}><\widetilde{f_n}|e^{-\beta
E_n^{(0)}}\sim \sum_n|E_n^{(0)}><E_n^{(0)}|e^{-\beta E_n^{(0)}}  \label{T.7}
\end{equation}
so the oscillators will be thermalized when $t\rightarrow \infty $, for
small interactions.

\section{Local and global time-asymmetry.}

Let us now precise the meaning of two important words: ''conventional'' and
''substantial'':

-In mathematics we use to work with identical objects, like points, the two
directions of an axis, the two semicones of a light cone, the two time
orientations of a time-oriented manifold, etc.

-In physics there are also identical objects, like identical particles, the
two spin directions, the two minima of a typical ''two minima'' potential,
etc.

-When (\cite{Sachs}, \cite{Penrose}) we are forced to call two identical
objects by different names we will say that we are establishing a {\it %
conventional difference}, e.g.: when we call $e_1$ and $e_2$ two electrons,
or ''up'' and ''down'' two spin directions, or ''right'' and ''left'' two
minima of a symmetric potential curve, while

- if we call two different objects by different names we will say that we
are establishing a {\it substantial difference.}

The problem of time asymmetry is that, in all time-symmetric normal physical
theories, usually the difference between past and future is just
conventional. In fact, if we can change the word ''past'' by the word
''future'', in these theories, the theory remains valid and nothing actually
changes. But we have the clear psychological filling that the past is
substantially different from the future. Thus the problem of the arrow of
time is to find theories where the past is substantially different from the
future, such that the usual well established physics would remain valid. As
we will see our minimal irreversible quantum mechanics would be one of these
theories.

But, up to now, we have just postulated that the states $\rho \in \Phi _{-}$
and the observables $R\in \Phi _{+},$ namely we have established a local
time asymmetry within a particular quantum system . To establish a global
time asymmetry for the whole universe is quite a different task.

It is clear that if we content ourselves with the local time asymmetry we
must face several problems. E. g. as $\Phi _{-\text{ }}$and $\Phi _{+}$ are
only conventionally different there is no reason to justify that all the
states would belong to a space $\Phi _{-}$ and all the operators to a space $%
\Phi _{+}$ in all the systems of the universe. Most likely the states would
belong to $\Phi _{-\text{ }}$ in the 50\% of the system of the universe and
to $\Phi _{+}$ in the other 50\%. For the operators we would have the
inverted situation. Then: what would happen when a system of the first class
interacts with a system of the second class? Perhaps somehow would an
average arrow of time be established?

To scape from this dilemma it is clear that we need a cosmological model
that correlates all the local arrows of time, and we know that, nowadays,
cosmological models are never completely satisfactory. Nevertheless the more
satisfactory, realistic, and simplest model to solve the problem is
Reichenbach global system \cite{Reichenbach} that we will further explain.

\section{Reichenbach global system.}

The set of {\it irreversible} processes within the universe, each one
beginning in an unstable non-equilibrium state, can be considered a{\it \
global system }\cite{Reichenbach}{\it , }\cite{Davies}{\it . }Namely, every
one of these branching processes began in a non-equilibrium state, such
that, this state was produced by a previous process of the set. E. g.: Gibbs
ink drop (initial unstable state) spreading in a glass of water (
irreversible process) it only may exist if it was produced by an ink factory
(since the probability to concentrate an ink drop by a fluctuation in a
glass, where the ink is nixed with the water, is extremely small). This
factory extract the necessary energy from an oven, where coal (initial
unstable state) is burning (branch irreversible process); in turn coal was
created with energy coming from the sun, where $H$ (initial unstable state)
is burning (branch irreversible process); finally $H$ was created using
energy obtained from the unstable initial state of the universe (the
absolute initial state of the global system). It is now clear that within
the global system all the arrows of time point to the same direction since
all of them are originated in the same global initial unstable state (why
the initial state is unstable is explained in papers \cite{Davies} and \cite
{Aquilano} and it is not of our concern in this paper).

Of course Reichenbach global system was originally imagined as a model of
the classical universe. But we can also imagine a quantum global system.

Let us draw the ordinary diagram of a scattering experiment (fig. 2) to have
a graphic idea of the nature of the unstable states. In the center of the
diagram there is a black box that symbolizes any scattering process. A set
of stables ''in'' states $a_1,a_2,...$ is transformed by the scattering
process in another set of stable ''out'' states $b_1,b_2,...$. It is a
reversible process because the evolution equations are time-reversible, so
we can interchange the ''in'' and ''out'' states and all the results remain
valid. In fact, fig. 2 is essentially symmetric. The variation of the
quantum entropy vanishes in this process .

Now, let us cut the black box in two parts by the dotted line draw at $t=0$.
Then, we can consider the right side of the figure, namely fig. 3. This
figure was introduced by A. Bohm \cite{Bohm}, so we will call this kind of
figures ''Bohm diagrams''. In fig. 3 the set of stable ''incoming'' states
creates a set of unstable states, $u_1,u_2,...$ which are growing states and
they belong to space $\phi _{+}$ \cite{L&P}$.$ (e. g. radiation exciting an
electron of the ground state). As the states of $\phi _{+}$ are linear
combinations of the states of $\phi _{-}^{\times }$, in some sense they can
also be considered also as growing states and as such they can be symbolized
as horizontal lines inside the half box. Fig. 3 is asymmetric and it
symbolizes an irreversible creation process. The evolution equation is still
time-symmetric but irreversibility is introduced by the growing nature of
the states of space $\phi _{-}^{\times }$ or by the non-invertible time
evolution operator acting on $\phi _{+}$ for $t<0$ . The variation of the
entropy is negative in this process. This is not contradictory since in
every creation process there is an incoming energy and then we can consider
the system as an open one.

We can also consider the second half of figure 1, namely fig. 4. It is the
Bohm diagram of a decaying process where a set of unstable decaying states $%
u_1,u_2,...,$ that are linear combinations of the basis of space $\phi
_{+}^{\times },$ is transformed in a set of stable ''outgoing'' stable
states of $\phi _{-}$ (e. g.: an excited electron decaying into the ground
state and emitting radiation). Fig 4 is asymmetric and symbolizes a decaying
irreversible process. Again, evolution equations are still time-symmetric
but the decaying nature of the states of space $\phi _{+}^{\times }$
introduces the irreversibility, etc., etc. The variation of the entropy is
positive in this process. These would be the diagrams corresponding to local
processes.

But Bohm diagrams allow us to also see also quantum structure of a global
system ( fig. 5). The universe is represented by a set of branching
scattering processes with one initial unstable state symbolized by the cut
box (at ''big-bang'' time $t=0)$ in the far right. Each subsystem going from
an unstable state to equilibrium (the ink drop spreading in the water, the
sugar lump solving in the coffee,...) is symbolized by a decaying process as
the one of fig 4, namely the diagram in the shaded box of figure 5. The
creation of an unstable state is symbolized by a creation process (like the
one of fig. 3) where energy comes from a previous decaying process (the ink
factory with its oven). One of these larger subsystem with its source of
energy is represented in the dotted box in fig. 5 . The overall process is
irreversible, because fig. 5 is asymmetric, and if we would have a model of
this universe ( see some simple models in ref.\cite{Lombardo}, \cite{Cosmo})
the state of the universe must belong to some global space $\phi _{-}^G$ or $%
\Phi _{-}^G.$ Therefore in this diagram there is a clear arrow of time. But
in the previous diagrams (fig. 3 or fig. 4) the arrow of time was a
''local'' one, while this diagram has one of the most important
characteristic of the observed time asymmetry: it is global. This is the way
to introduce the arrow of time in a time-symmetric dynamical formalism: by a
global and generalized symmetry breaking process

But we must remember that the difference between the global $\Phi _{-}^G$
and the global $\Phi _{+}^G$ of the whole universe global system is just
conventional since these two spaces are identical. Thus physics is the same
in $\Phi _{-}^G$ than in $\Phi _{+}^G.$ Think in a cosmological model, (fig.
5) life will be the same in this universe (with a quantum state in space $%
\Phi _{-}^G)$ than in the universe of fig. 6, the time inverted image of
fig. 5, (with a t-inverted quantum state in space $\Phi _{+}^G).$ In fact,
since in both models of the universe (if completely computed) all the arrows
of time must point to the same direction, there is no physical way to decide
if we are in one model or in the other. So both models are identical. Thus
the choice between $\Phi _{-}^G$ and $\Phi _{+}^G$ is just conventional and
physically irrelevant (as to choose one of the two minima of the potential
in spontaneous symmetry breaking problem).

But once this choice is made a substantial difference is established in the
model e. g. the only time evolution operator is $U_{-}(t)=e^{-iHt},t>0,$ and
it cannot be inverted, we only have equilibrium towards the future, entropy
growing, etc. .

Once the $\Phi _{-}^G$ or the $\Phi _{+}^G$ is chosen in the global system
we are forced to choose the corresponding spaces in the local subsystems, if
we want to study these subsystems as isolated systems, and a global arrow of
time is established. This fact solves the problem stated in the preceding
section.

Thus the choice between $\Phi _{-}^G$ and $\Phi _{+}^G$ is trivial and
unimportant (but this choice must be a global one), that is why the arrow of
time is not introduced by hand in our theory. The important choice is
between ${\cal L}^G$ and $\Phi _{-}^G$ (or $\Phi _{+}^G)$ as the space of
our physical admissible states. And we are free to make this choice, since a
good physical theory begins by the choice of the best mathematical structure
to mimic real nature. {\it Thus, our thesis is essentially that time
asymmetric mathematical structures like ours mimic in the more economical
way the time-asymmetric of the Nature where we live, than time-symmetric
mathematical structures.}

\section{Other results.}

The main results related with quantum mechanics are stated in the above
sections. But we must comment that using the present formalism all the
relevant results of irreversible statistical mechanics can also be obtained,
e. g. all the results of the book \cite{Balescu}, as it is proved in ref. 
\cite{CGG}, because the main $\Pi $ projector of the quoted book can be
defined using Gel'fand triplets. Also, in some simple cases, we can go from
the quantum models to the classical ones \cite{Diener}, where we find the
same philosophy, in classical cases. Chaotic models like Baker's
transformation and Renyi's maps, are also treated with the same method, with
good results \cite{Hasegawa}, \cite{Antonioutasaki}. Other interesting
results are contained in papers \cite{Bohm}, \cite{Sudarshan}, \cite{Chiu}, 
\cite{Antoniou}, and \cite{QAT}. So what we have explained is just the
quantum axiomatic formalism of a general method to deal with irreversible
processes.

\section{Conclusion.}

We claim that our axiomatic formalism condenses the main ideas of how to
solve the problem of time asymmetry pioneered by many authors that appears
in the bibliographical references. This method is based in the reasonings we
have made in the introduction and it yields correct physical results that
coincide with those obtained by other methods (coarse-graining, traces,
stochastic noises, lost of information, etc.).

We do not foresee any cross-experiment to settle which of the quoted methods
(coarse-graining, traces, stochastic noises, lost of information, etc.) or
ours is the correct one, because we think that they are somehow
complementary and that they only explain the real physical world in
different ways. Therefore we believe that we must solve at least three
problems in order to complete our theory:

a.- The methods of coarse-graining: traces, projections, stochastic noises,
lost of information, etc. have clear physical motivations. On the contrary
our method is just based on the search of the minimal mathematical structure
to explain time-irreversibility. Even if this reason can be a sufficient one
for mathematical minded readers, it turns out to be not so eloquent for
physical minded ones. So we must find the relation among all the method
because most probably they are all based in the same or very similar
(eloquent) physical bases. This work was already began in paper \cite{mex}
where it is shown that causality is the real reason for the choice of the
proper space of physical states.

b.-We have admitted, under eq. (\ref{2.7}), that there is not a unique way
to define space $\phi _{-}.$ We must find the necessary and sufficient
condition to define a unique space $\phi _{-}$ or at least a class of spaces
all of them endowed with enough properties to explain all phenomena related
with time-asymmetry. Lax-Phillips scattering theory, recently redeveloped by
Boris Pavlov \cite{Pavlov}, and the quoted paper \cite{mex} seem the more
promising ways to solve the problem.

c.- Essentially space $\Phi _{-}$ is the one where all out-of-equilibrium
entropies turn out to be growing. But we have said, at the beginning of
section 13, that a complete study that single out, in an indisputable way,
an entropy, that generalizes the phenomenological thermodynamical entropy at
equilibrium, is missing. Most likely only when this task will be
accomplished we will be able to motivate the choice of space $\Phi _{-}$
convincingly$.$

Meanwhile we just claim that our formalism is a contribution to the
understanding of time-asymmetry.

\section{Acknowledgment.}

This work was partially supported by grants: CI1$^{*}$-CT94-0004 and PSS$%
^{*} $-0992 of the European Community, PID-0150 of CONICET (National
Research Council of Argentina), EX-198 of the Buenos Aires University, and
12217/1 of Fundaci\'{o}n Antorchas and also grants from the Fondation pour
la Receherche Fondamentale OLAM and the British Council.

\section{Appendix . Wigner function integral and classical entropy.}

Let $\rho $ be a density matrix of Liouville space ${\cal L}={\cal H\oplus
(H\otimes H)}$ and let $\{|q>\}$ be the configuration or position basis of
the Hilbert space ${\cal H}.$ The Wigner function corresponding to state $%
\rho $ is a real but not positive definite quantity that reads (\cite{Wigner}
eq. (2.1)): 
\begin{equation}
\rho _W(q,p)=\pi ^{-1}\int <q-\lambda |\rho |q+\lambda >e^{2i\lambda
p}d\lambda  \label{B.1}
\end{equation}
This equation is valid if the wave function has only one variable. If it
would have $n$ variables the $\pi ^{-1}$ must be changed by $\pi ^{-n}.$ It
can be proved that: 
\begin{equation}
L\rho _W(q,p)=\pi ^{-1}\int <q-\lambda |L\rho |q+\lambda >e^{2i\lambda
p}d\lambda +O(\hbar )  \label{B.2}
\end{equation}
where $L$ is respectively the classical and quantum Liouville operator. In
the classical limit $\hbar \rightarrow 0$ therefore $\rho _W$ can be
considered as the classical distribution function corresponding to $\rho $.
As in the classical regime we practically work in this limit we will
consider that eq.(\ref{B.1}) is the relation between the quantum density
matrix and the classical distribution function. In fact, even if $\rho _W$
is not generally positive definite, using the Wigner integral from classical
equations we can obtain quantum equations and vice versa, as we will see in
a few examples. E.g., let us observe that (\cite{Wigner} eq. (2.6): 
\begin{equation}
\parallel \rho _W\parallel =\int \int \rho _W(q,p)dqdp=  \label{B.3}
\end{equation}
\[
=\int dq\int <q-\lambda |\rho |q+\lambda >\delta (\lambda )d\lambda =Tr\rho 
\]
so to the classical norm corresponds the quantum trace. Also, if we define
the classical analogue of operator $O$ as (\cite{Wigner} eq. (2.12)): 
\begin{equation}
O_W(q,p)=\int dze^{ipz}<q-\frac 12z|O|q+\frac 12z>  \label{B.3'}
\end{equation}
we obtain that : 
\begin{equation}
(\rho _W|O_W)=\int \int \rho _W(q,p)O_W(q,p)dpdq=Tr(\rho O)  \label{B.4}
\end{equation}
Therefore to the inner product in classical Liouville space corresponds the
inner product in the quantum Liouville space. Finally if $O_1$ and $O_2$ are
two operators and $O_P=O_1O_2$ it is ( \cite{Wigner} eq. (2.59)): 
\begin{equation}
O_{PW}(q,p)=O_{1W}(q,p)e^{\frac \Lambda {2i}}O_{2W}(q,p)  \label{B.5}
\end{equation}
where: 
\begin{equation}
\Lambda =\overleftarrow{\frac \partial {\partial p}}\overrightarrow{\frac %
\partial {\partial q}}-\overleftarrow{\frac \partial {\partial q}}%
\overrightarrow{\frac \partial {\partial p}}  \label{B.6}
\end{equation}
Therefore if $O_1=O_2$ we have that: 
\begin{equation}
O_{PW}(q,p)=[O_{1W}(q,p)]^2  \label{B.7}
\end{equation}
This fact completes the analogy between classical and quantum spaces
implemented by the Wigner integral.

As Wigner integral (\ref{B.1}) is a linear mapping the quantum evolution
equation (\ref{12.2}) can be reproduced in the ''classical case''as 
\begin{equation}
\rho _W(t)=\rho _{*W}+e^{-\gamma _{\min }t}\rho _{1W}(t)  \label{B.8}
\end{equation}
where, somehow, $\rho _W$ could be considered as a classical distribution
function. It is not so because it is not positive definite. But we will see
that $\rho _W$ has this property near equilibrium, where we know that, due
to the appearance of decoherence and correlations, the state is classic.

In fact, the classical states are defined as those having decoherence and
correlations (this fact can be seem in ref. \cite{Zurek}, \cite{Juanpa}
using coarse-graining method or using our method, in section 11 of this
paper for decoherence and in ref. \cite{Lombardo} for correlations), namely
those that can be expanded as: 
\begin{equation}
\rho =\sum_I\rho _I|I>_{cor}<I|_{cor},\text{ }\rho _I\geq 0  \label{12.13}
\end{equation}
where $|I>_{cor}$ is a no-ghost state such that its position $<q|I>_{cor}$
and its momentum $<p|I>_{cor}$ are correlated, i. e. they are as defined as
possible (e. g. as the ground state of a H atom). Precisely, they are as
defined as it is allowed by the uncertainty principle, around a point $%
(q_I,p_I)$. The quantum equilibrium state $\rho _{*}$ is one of these
states. Let $\rho _{*W\text{ }}$ be the classical equilibrium states . The
Wigner integral of this state gives: 
\begin{equation}
\rho _{*W}(q,p)=\pi ^{-1}\int_{-\infty }^{+\infty }<q-\lambda |\rho
_{*}|q+\lambda >e^{-2i\lambda p}d\lambda =  \label{12.14}
\end{equation}
\[
=\pi ^{-1}\sum_I\rho _{I*}\int_{-\infty }^{+\infty }<q-\lambda
|I>_{cor}<I|_{cor}|q+\lambda >e^{-2i\lambda p}d\lambda = 
\]
\[
\cong \pi ^{-1}\sum_I\rho _{I*}\int_{-\infty }^{+\infty }<q_I-\lambda
|I>_{cor}<I|_{cor}|q_I+\lambda >\delta (\lambda )e^{-2i\lambda p}d\lambda = 
\]
\[
=\pi ^{-1}\sum_I\rho _{I*}\int_{-\infty }^{+\infty
}|<q_I|I>_{cor}|^2e^{-2i\lambda p}\delta (\lambda )d\lambda =\pi
^{-1}\sum_I\rho _{I*}|<q_I|I>_{cor}|^2\geq 0 
\]
since the functions $<q_I|I>_{cor},$ $<I|_{cor}|q_I>$ vanish except around a
definite value of $q_I,$ so the states $<q\pm \lambda |$ are only different
from zero when $q\cong q_I$ and $\lambda \simeq 0,($this is the reason of
the factor $\delta (\lambda ))$ and because of eq. (\ref{6.4}) which is
valid for $\rho _{*}.$ Then we conclude that the equilibrium states and the
quantum states near $(q_I,p_I)$, have the property $\rho _W(q,p)\geq 0$ and
therefore these Wigner functions can be considered classical states. These
states disappear, if we wait long enough, due to the dumping term in eq. (%
\ref{B.8})

So, in this classical regime, a usual conditional entropy can be defined,
and any quantum formula can be reobtained in the classical case. Of course
we can directly work in this regime if we consider the classical analogue of
our formalism and the poles of the classical Liouville operator \cite
{Hasegawa}. But now we know, from papers \cite{Mackey} and \cite{Voigt},
that this classical conditional entropy never decreases. So we can consider $%
\rho _{W\text{ }}$ and make a projection obtaining $\widetilde{\rho }_W$ and
define: 
\begin{equation}
S(\widetilde{\rho }_W)=\int_\Gamma \widetilde{\rho }_W\log \frac{\widetilde{%
\rho }_W}{\rho _{*W}}dqdp  \label{B.9}
\end{equation}
As this entropy is never decreasing we know that $\stackrel{\bullet }{S(%
\widetilde{\rho }_W)}\geq 0.$ Also as $\rho _W$ is given by the sum of local
terms (\ref{12.13}) this definition contains the notion of local equilibrium
entropy. Now we can repeat everything we have done in the quantum case,
since condition 1, 2, and 3 can be translated to the quantum case.
Precisely: Condition 1 using eq. (\ref{B.3}). Condition 2 using eq. (\ref
{B.5}). Condition 3 using eq. (\ref{B.1}). So we obtain: 
\begin{equation}
S(\widetilde{\rho }_W)=-\frac 12e^{-2\gamma _{\min }t}\int_\Gamma \frac{%
\widetilde{\rho }_{1W}^2}{\rho _{*W}}dqdp+...  \label{B.10}
\end{equation}
where the dots symbolize higher order terms. Thus: 
\begin{equation}
\stackrel{\bullet }{S}(\widetilde{\rho }_W)=\gamma _{\min }e^{-2\gamma
_{\min }t}\int_\Gamma \frac{\widetilde{\rho }_{1W}^2}{\rho _{*W}}dqdp+...
\label{B.11}
\end{equation}
So now, it is clear that $\stackrel{\bullet }{S}(\widetilde{\rho }_W)>0$ and
that we have obtained the second law of the thermodynamics. This is not at
all surprising since it is well known that, by a projection, the constant $%
S(\rho _W)$ can be transformed in the variable $S(\widetilde{\rho }_W).$ We
have just showed how our method works and also how we can obtain the
classical eqs. (\ref{B.8}) and (\ref{B.10}).

So our $S(\widetilde{\rho }_W)$ has all the properties that we have
announced and we can claim that it is a good candidate to play the role of
internal entropy.

\section{Figures.}

Fig 1. Plane $\nu $ and curve $C.$

Fig 2. A scattering experiment.

Fig. 3. The left hand side of fig 1: a growing process.

Fig. 4. The right hand side of fig. 1: de decaying process.

Fig. 5. The quantum branch system of the universe.

Fig. 6. Fig. 4 time-inverted.


\begin{references}
\bibitem{Messiah}  Messiah A. {\it Quantum mechanics,} North-Holland Co.
Amsterdam, 1961.

\bibitem{CGG}  Castagnino M., Gaioli F., Gunzig E., Found. Cos. Phys., {\bf %
16,} 221, 1996.

\bibitem{Sachs}  Sachs R. G., {\it The physics of time reversal,} Univ. of
Chicago Press, Chicago, 1987.

\bibitem{Feynman}  Feynman R. P., Leighton R. B., Sand M., {\it Lectures on
physics,} Addison-Wesley Pub. Co., Reading, 1963.

\bibitem{Bohm}  Bohm A., {\it Quantum Mechanics: Foundations and
Applications,} Springer-Verlag, Berlin, 1986.

\bibitem{Ballentine}  Ballentine L., {\it Quantum Mechanics,}Prentice Hall,
Englewoods Cliffs, 1990.{\it \ .}

\bibitem{L&P}  Lax P. D.,Phillips R. S., {\it Scattering theory,} Acad.
Press, New York, 1967.

\bibitem{AntoniouBohm}  Antoniou I.,Bohm A., Kielanowski P., Jour. Math
Phys. {\bf 36,} 1, 1995.

\bibitem{Bohm51}  Bohm A. Phys. Rev., {\bf 51,} 1758, 1995.

\bibitem{mex}  Castagnino M., {\it Rigged Hilbert spaces, duality cosmology,}
Proceeding First Latin American Symposium on High Energy Physics, Ed.
D'Olivo J. C. et al. M\'{e}rida, Mexico, 1996, p. 469, AIP Conference
Proceedings 400, Woodbury, New York, 1997.

\bibitem{Diener}  Castagnino M., Diener R., Lara L., Puccini G.,{\it \
Instability and time-asymmetry: the case of the up side down harmonic
oscillator,} Int. Jour. Theo. Phys, in press, 1997.

\bibitem{Gadella}  Castagnino M., Gadella M., Gaioli F., Laura R., {\it %
Gamov vectors and time-asymmetry, }submitted to Fortschrift der Physik, 1997.

\bibitem{Castscape.}  Castagnino M., Gunzig E., {\it A landscape on
time-asymmetry, }Int. Jour. Theo. Phys., in press, 1997.

\bibitem{Trio}  Castagnino M., Domenech G., Levinas M. L., Umerez N., Jour
Math. Phys. {\bf 37,} 2107, 1996.

\bibitem{Laucast}  Castagnino M., Laura R. Phys. Rev. A. {\bf 56,} 108, 1996.

\bibitem{Sudarshan}  Sudarshan E. C. G., Chiu C. B., Gorini V., Phys. Rev.
D, {\bf 18, }2914, 1978

\bibitem{Sudarshanu}  Sudarshan E. C. G. Phys. Rev. A, {\bf 46,} 37, 1992.

\bibitem{Roberto}  Laura R., {\it Scattering and intrinsic irreversibility,}
Int. Journ. Theo. Phys, in press 1997.

\bibitem{Nakanishi}  Nakanishi N., Prog. Theo. Phys., {\bf 19,} 607, 1958.

\bibitem{Mackey}  Mackey M. C., Rev. Mod. Phys., {\bf 61,} 981, 1989.

\bibitem{Chiu}  Parravicini G.,, Gorini V., Sudarshan E. C. G., J. Math
Phys., {\bf 21,} 2208, 1980..

\bibitem{Jordan}  Jordan T. F., {\it Linear operator for quantum mechanics, }%
Wiley, New York, 1969.

\bibitem{Juanpa}  Hu B. L., Paz J. P., Zhang Y., Phys. Rev. D, {\bf 45,}
2843, 1992.

\bibitem{LaucastIII}  Laura R.,Castagnino M., {\it A minimal irreversible
quantum mechanics: the mixed states and the diagonal singularity}, Phys.
Rev. A, in press,1997

\bibitem{Lombardo}  Castagnino M., Lombardo F., Gunzig E., Gen. Rel. Grav., 
{\bf 27, }257, 1995.

Castagnino M., Lombardo F., Gen. Rel. Grav., {\bf 28,} 263, 1996.

Castagnino M., {\it the mathematical structure of quantum superspace as a
consequence of time-asymmetry, }Phys. Rev. D., in press, 1997.

\bibitem{Gell'man}  Zurek W. H., {\it Preferred sets of states,
predictability, classicality, and environment induced decoherence, }in:
Physical Origin of Time Asymmetry, p.175, Halliwell et al. ed. Cambridge
Univ. Press, Cambridge, 1994.

\bibitem{Arbo}  Arb\'{o} D., Castagnino M., Gaioli F., Iguri S., {\it %
Minimal irreversible quantum mechanics: the brownian motions, }in
preparation 1997.

\bibitem{Leibowitz}  Lebowitz J. L., {\it Time's arrow and Boltzmann
entropy, } Physical Origin of Time Asymmetry, p. 131, Halliwell et al. eds.
Cambridge Univ. Press, Cambridge, 1994.

\bibitem{Polemica}  Spohn H. J. Math Phys., {\bf 19,} 1227, 1978.

\bibitem{Zurek}  Zurek W. H., Physics Today, {\bf 44,} 36, 1991.

\bibitem{Hasegawa}  Hasegawa H. H., Dreiden D. J., Phys. Rev. E, {\bf 50,}
1781, 1994.

\bibitem{Penrose}  Penrose R.,{\it \ Singularities and time asymmetry,} in%
{\it \ General Relativity, an Einstein centenary survey,} Ed. Hawking S.,
Israel W., 1979.

\bibitem{Reichenbach}  Reichenbach H., {\it The direction of time,} Univ. of
California Press, Berkeley, 1956.

\bibitem{Davies}  Davies P.,{\it \ Stirring up trouble, }in Physical Origin
of Time-asymmetry, Halliwell J. J. et al. eds., Cambridge Univ. Press,
Cambridge, 1994.{\it \ }

\bibitem{Aquilano}  Aquilano R., Castagnino M., Mod. Phys. Lett. A., {\bf 11,%
} 755, 1996.

Aquilano R., Castagnino M., Astroph. Space Sci. {\bf 238,} 159, 1996.

\bibitem{Cosmo}  Castagnino M., Gaioli F., Sforza D., Gen. Rel. Grav., {\bf %
28,} 1129, 1996.

Castagnino M., Gunzig E., Nardone P., Prigogine I., Tasaki S., {\it Quantum
cosmology and large Poincar\'{e} systems,} in Quantum, Chaos, and Cosmology,
p. 3, Namiki M. ed., AIP books, Woodbury, New York, 1996.

\bibitem{Balescu}  Balescu R., {\it Equilibrium and non-equilibrium
statistical mechanics, }J. Wiley \& Sons, New York, 1963.

\bibitem{Antonioutasaki}  Antoniou I., Tasaki S., Physica A, {\bf 190,} 303,
1991,

Antoniou I., Tasaki S., Int. J. Quant. Chem., {\bf 46, }427, 1993.

\bibitem{Antoniou}  Antoniou I., Prigogine I., Physica A, {\bf 192,} 443,
1993.

\bibitem{QAT}  Bohm A., Loewe M., Maxson S., {\it Microphysical
irreversibility in quantum mechanics, }Reports on Theo. Phys., in pres, 1996.

\bibitem{Wigner}  Hillery M., O'Connell R. F., Scully M. D., Wigner E. P.,
Phys. Rep. {\bf 106}, 123, 1984.

\bibitem{Pavlov}  Pavlov B. Proceeding Actual Problems in Quantum Mechanics
Seminar, Peyresq, 1997, Int. Jour. Theo. Phys., in press, 1997.

\bibitem{Voigt}  Voigt J., Comm. Math. Phys., {\bf 81,} 31, 1981.
\end{references}
\end{document}